%% file: main.tex
\documentclass[twocolumn,twocolappendix]{aastex631}

\usepackage{soul}

\newcommand{\vmon}{V960$\,$Mon}
\newcommand{\vmonE}{V960$\,$Mon$\,$E}
\newcommand{\vmonSE}{V960$\,$Mon$\,$SE}
\newcommand{\vmonN}{V960$\,$Mon$\,$N}
\newcommand{\vmonNE}{V960$\,$Mon$\,$NE}
\newcommand{\vmonS}{V960$\,$Mon$\,$S}
\newcommand{\bsix}{band$\,$6}
\newcommand{\bfour}{band$\,$4}
\newcommand{\bthree}{band$\,$3}
\newcommand{\kms}{$\,$km$\,$s$^{-1}$}
\newcommand{\ceo}{C$^{18}$O}
\newcommand{\ntdp}{N$_2$D$^+$}
\newcommand{\dcop}{DCO$^+$}
\newcommand{\sio}{SiO}

\include{acknowledgments}

\shorttitle{The large-scale view of \vmon{}}
\shortauthors{Weber et al.}

\begin{document}

\title{A Multi-Wavelength Study of the Dynamic Environment surrounding the FUor V960~Mon}

\author[0000-0002-3354-6654]{Philipp Weber}
\email{philipppweber@gmail.com}
\affiliation{Departamento de Física, Universidad de Santiago de Chile, Av. Victor Jara 3659, Santiago, Chile}
\affiliation{Millennium Nucleus on Young Exoplanets and their Moons (YEMS), Chile}
\affiliation{Center for Interdisciplinary Research in Astrophysics and Space Exploration (CIRAS), Universidad de Santiago de Chile, Chile}
\author[0000-0002-9966-9403]{Silvio Ulloa}
\affiliation{Departamento de Astronomía, Universidad de Chile, Casilla 36-D, Santiago, Chile}
\affiliation{Millennium Nucleus on Young Exoplanets and their Moons (YEMS), Chile}
\author[0000-0003-2953-755X]{Sebastián Pérez}
\affiliation{Departamento de Física, Universidad de Santiago de Chile, Av. Victor Jara 3659, Santiago, Chile}
\affiliation{Millennium Nucleus on Young Exoplanets and their Moons (YEMS), Chile}
\affiliation{Center for Interdisciplinary Research in Astrophysics and Space Exploration (CIRAS), Universidad de Santiago de Chile, Chile}
\author[0000-0002-1575-680X]{James Miley}
\affiliation{Departamento de Física, Universidad de Santiago de Chile, Av. Victor Jara 3659, Santiago, Chile}
\affiliation{Millennium Nucleus on Young Exoplanets and their Moons (YEMS), Chile}
\affiliation{Center for Interdisciplinary Research in Astrophysics and Space Exploration (CIRAS), Universidad de Santiago de Chile, Chile}
\author[0000-0002-2828-1153]{Lucas Cieza}
\affiliation{Instituto de Estudios Astrofisicos, Facultad de Ingeniería y Ciencias, Universidad Diego Portales, Av. Ejercito 441, Santiago, Chile}
\affiliation{Millennium Nucleus on Young Exoplanets and their Moons (YEMS), Chile}
\author[0000-0002-6166-2206]{Sergei Nayakshin}\affiliation{School of Physics and Astronomy, University of Leicester, Leicester, LE1 7RH, UK}
\author[0000-0002-5903-8316]{Alice Zurlo}
\affiliation{Instituto de Estudios Astrofisicos, Facultad de Ingeniería y Ciencias, Universidad Diego Portales, Av. Ejercito 441, Santiago, Chile}
\affiliation{Millennium Nucleus on Young Exoplanets and their Moons (YEMS), Chile}\affiliation{Escuela de Ingeniería Industrial, Facultad de Ingeniería y Ciencias, Universidad Diego Portales, Av. Ejercito 441, Santiago, Chile}
\author[0000-0003-2300-2626]{Hauyu Baobab Liu}
\affiliation{Department of Physics, National Sun Yat-Sen University, No. 70, Lien-Hai Road, Kaohsiung City 80424, Taiwan, R.O.C.}
\affiliation{Center of Astronomy and Gravitation, National Taiwan Normal University, Taipei 116, Taiwan}
\author[0000-0002-4283-2185]{Fernando Cruz-Sáenz de Miera}\affiliation{Institut de Recherche en Astrophysique et Plan\'etologie, Universit\'e
de Toulouse, UT3-PS, OMP, CNRS}
\author[0000-0001-5073-2849]{Antonio Hales}
\affiliation{8 National Radio Astronomy Observatory, 520 Edgemont Rd., Charlottesville, VA 22903-2475, USA}
\author[0000-0002-4266-0643]{Antonio Garufi}\affiliation{INAF – Istituto di Radioastronomia, Via Gobetti 101, 40129 Bologna, Italy}
\author[0000-0002-4502-8344]{Dimitris Stamatellos}\affiliation{Jeremiah Horrocks Institute for Mathematics, Physics and Astronomy, University of Central Lancashire, Preston PR1 2HE, UK}
\author[0000-0001-7157-6275]{Ágnes Kóspál}\affiliation{Konkoly Observatory, HUN-REN Research Centre for Astronomy and Earth Sciences, MTA Centre of Excellence, Konkoly-Thege Mikl\'os \'ut 15-17, 1121 Budapest, Hungary}
\affiliation{Institute of Physics and Astronomy, ELTE E\"otv\"os Lor\'and University, P\'azm\'any P\'eter s\'et\'any 1/A, 1117 Budapest, Hungary}
\affiliation{Max-PLanck-Insitut f\"ur Astronomie, K\"onigstuhl 17, 69117 Heidelberg, Germany}
\author[0000-0003-4784-3040]{Viviana Guzmán}
\affiliation{Instituto de Astrofísica, Pontificia Universidad Católica de Chile, Av. Vicuña Mackenna 4860, 7820436 Macul, Santiago, Chile}
\affiliation{Millennium Nucleus on Young Exoplanets and their Moons (YEMS), Chile}

\begin{abstract}
The evolution of young stars and planet-forming environments is intrinsically linked to their nascent surroundings. This is particularly evident for FU Orionis (FUor) objects—a class of young protostars known for dramatic outbursts resulting in significant increases in brightness.
We present a case study of \vmon{}, an FUor that has recently been found to show signs of a fragmenting spiral arm, potentially connected to planet formation.
Our study explores the large-scale environment ($10^3-10^4$\,au) and incorporates ALMA \bthree{}, \bfour{} and \bsix{} continuum data, molecular emissions from $^{12}$CO, $^{13}$CO, \ceo, \sio{}, \dcop, \ntdp and DCN, alongside optical and near-infrared observations from VLT/MUSE and VLT/SPHERE.
We map a region of $20''$ across where we find tantalizing emissions that provide a unique view of a young group of protostars, including the discovery of a class-0 protostar to the east of the FUor.
The $^{12}$CO and SiO tracers suggest that this object is at the base of an outflow, potentially impacting the surrounding medium.
The MUSE and SPHERE observations indicate the presence of an elongated feature towards a prominent source to the southeast that may represent interaction between \vmon{} and its surrounding. 
Moreover, the \ceo{} emission overlaps with the clumps of the detected fragmenting spiral arm.
These findings provide the strongest evidence to date for a connection between infalling material, fragmentation, and the intensity outburst of a protostar.
Our case study highlights the complex interactions between young stars and their surroundings that drive the evolution of the planet forming environment.
\end{abstract}

\section{Introduction} \label{sec:intro}
FUor events represent a transient phase in early stellar evolution, characterized by sudden and dramatic increases in luminosity. These luminous outbursts, typically elevating the star's brightness by several magnitudes, are believed to result from episodic accretion events, during which a significant amount of material from the circumstellar disk is rapidly dumped onto the star \citep{Hartmann1996, Audard2014}.
Such events not only alter the star's appearance but also have profound implications for the evolution of the surrounding protoplanetary disk.
It has been suggested that these episodic accretion events could play a pivotal role in the early stages of planet formation.

Firstly, the occurrence of an accretion burst has been suggested to be directly linked to planet-forming processes such as the Gravitational Instability (GI) \citep{Armitage2001,Zhu2009,Stamatellos2011}\footnote{For a broader discussion on the origin of FUor events, see the review by \citet{Fischer2023}}. This association makes FUor objects prime targets for searching for signs of planet formation via GI. An important question in this context is how the high disk mass required to trigger GI can be sustained.

Secondly, the feedback from the enhanced accretion rates and consequent disk heating during FUor outbursts may influence the coagulation and composition of dust grains, potentially accelerating the formation of planetesimals \citep[e.g.][]{Zhu2009}. As such, understanding FUor phenomena could provide valuable insights into the dynamic processes underpinning the birth and early evolution of planetary systems.

Building on these theoretical foundations, observational studies have significantly advanced our understanding of FUor phenomena and their implications for planet formation.
Early observational efforts primarily focused on documenting the outburst characteristics, such as the rapid rise in brightness and spectral evolution \citep{Herbig1977, Hartmann1985}.
More recently, high-resolution observations in the (sub)millimetre and polarimetric imaging in the near-infrared (NIR) have provided more information on the FUors' surroundings.
Some of these observations have revealed intricate details of the circumstellar environment during and after FUor outbursts, including changes in disk structure, temperature profiles, and dust grain properties \citep{Cieza2016,Zurlo2017, Takami2018,Zurlo2024}.
The rapid temperature rise around FUor objects also reveals their chemical composition by evaporating disk compounds that would otherwise remain in solid form.
This has facilitated the study of disks' chemistry, in particular the detection and study of complex organic molecules (\citealp{Lee2019}, \citealp{Jeong2025}, Cruz-Sáenz de Meira et al. subm.) and water vapor \citep{Tobin2023}. 

Recent numerical simulations support the notion that the origin of an FUor event can trace back to the extensive environment of the young star.
For example, \citet{Kuffmeier2018} found that the material that falls from the molecular cloud onto the circumstellar disc could locally induce GI.
GI facilitates the rapid transport of angular momentum and a subsequent accretion burst.
Similarly, \citet{Dullemond2019} described a scenario in which material is captured from an adjacent cloudlet, triggering an accretion burst.

In a broader context, the structure and composition of the disk have been found to depend on the interaction with the environment, which can affect the processes of stellar evolution and planet formation \citep{Pineda2023,Winter2024}.

\vmon{} has received significant interest in the astronomical community due to its characteristic outburst properties and potential insights into early stellar evolution.
First identified in its outburst state in 2014 \citep{Maehara2014, Hillenbrand2014}, \vmon{} presents an exceptional opportunity to study the mechanisms and consequences of FUor-type eruptions in real time.
According to Gaia DR3, the system is located at a distance of $2189 \pm 281$\,pc \citep{GaiaDR3}. 
We employ this value with caution, as a high RUWE indicates possible systematic uncertainties, but it remains our distance estimate throughout this study.
At a distance of \(\gtrsim2\)\,kpc—which prevents resolving substructures around individual objects—this system offers an exceptional view into the accretion dynamics of young stellar systems and their interaction with the protostellar envelope. 

\vmon{} poses a spectacular circumstellar environment, which putatively revealed the first evidence of GI-fragments of planetary masses. 
\citet{Weber2023} presented a polarimetric image in the NIR obtained with the VLT/SPHERE instrument which shows characteristic large-scale spiral arms.
A complementing ALMA observation detected that one of those spiral arms seems to be undergoing fragmentation, indicated by several clumps of mm-emission that align along a spiral arm.
Here, we further study the case of \vmon{}, turning our attention to the environment at an even larger scale ($\sim$10$^{4}$\,au), presenting continuum and molecular line emission in the mm regime and scattered light in the optical and NIR.

In Section~\ref{sec:Observations} we describe the details of the observations and data reduction, we present main results in Section~\ref{sec:results} and discuss their implications in Section~\ref{sec:discussion}.
In Section~\ref{sec:conclusions}, we summarize our findings.

\section{Observations and Data Reduction}\label{sec:Observations}
\subsection{ALMA}
\vmon{} was observed by ALMA in \bsix{} (2016.1.00209.S, PI: M.Takami) on 20 April 2017 (compact configuration) and on 27 July 2017 (extended configuration).
The observation included in total five spectral windows: two of a bandwidth of 1.875\,GHz dedicated to the continuum at central frequencies of 216.877\,GHz and 232.178\,GHz; and three spectral windows of a bandwidth of 0.059\,GHz to cover the $J=2-1$ transitions of $^{12}$CO, $^{13}$CO and \ceo{}  centered at 230.514\,GHz, 220.375\,GHz and 219.537\,GHz, and channel widths of 15.259\,kHz, 30.158\,kHz, and 30.158\,kHz, respectively.

The \bsix{} continuum data of a reduced field of view (FOV) around \vmon{} were presented in \citet[][${\rm FOV}\sim4^"$]{Kospal2021} and further analyzed in \citet[][${\rm FOV}\sim2^"$]{Weber2023}.
Here, we used the \bsix{} continuum reduction with natural weighting presented in \citet{Weber2023}, where self-calibration was performed on the compact and extended configurations separately before combining them. 
For details about the data reduction, we refer the reader to that article.
For the spectral windows targeting the CO molecular lines, we subtracted the continuum from the lines and cleaned each channel with the CASA task {\tt tclean}.
We predefined cleaning masks by only permitting signal above three times the channel's rms-noise, but where appropriate, we added user-defined masks to avoid losing the faint signal that reappears over several channels.
We further inspected the continuum channels for the detection of molecular lines (see Appendix~\ref{appendix:spectrum}), and identified SiO, \ntdp{}, \dcop{} and DCN.
Because each molecular emission region has a unique spatial extent, we examined each case individually to identify baselines that produced sinusoidal artifacts from extended emission contamination and flagged them accordingly.

ALMA \bthree{} and \bfour{} data (both 2019.1.01144.S, PI: H.B.~Liu) were taken on the 23 August 2021 and 22 August 2021, respectively.
We reduced these data using natural weights.
Both observations included four spectral windows dedicated to the continuum of a bandwidth of 1.875$\,$GHz.
They were centered at 86.00\,GHz, 87.86\,GHz, 98.19\,GHz and 100.00\,GHz for \bthree{} and 146.00\,GHz, 147.86\,GHz, 158.19\,GHz and 160.00\,GHz for \bfour{}.
Because the continuum emission shows a low signal-to-noise ratio (SNR) in bands 3 and 4, and no significant improvement was expected, we did not perform any self-calibration.
We list the resulting dimensions of the clean beam and the SNR of each band in Table~\ref{tab:ALMA_obs}.

\begin{table}[]
    \centering
    \caption{List of archival ALMA observations. The measured rms-noise, $\sigma_{\rm rms}$, is given in µJy beam$^{-1}$.}
    \begin{tabular}{lccr}
        \tableline
       ALMA Band & Beam & $\sigma_{\rm rms}$ & Project-ID \\
       \tableline
       Band~6  & $0\farcs20\times0\farcs14$ & 28 & 2016.1.00209.S\\
       Band~4  & $0\farcs07\times0\farcs05$ & 25 & 2019.1.01144.S\\
       Band~3  & $0\farcs11\times0\farcs08$ & 23 & 2019.1.01144.S\\
       \tableline
    \end{tabular}
    \label{tab:ALMA_obs}
\end{table}

\subsection{VLT/MUSE}\label{subsec:data_MUSE}
\vmon{} was observed on the night of 24 January 2021 using the Multi Unit Spectroscopic Explorer (MUSE) at the VLT (0106.C-0510, PI: F.~Cruz-Saenz de Miera) with a total integration time of 25.6\,min.
The observing conditions were excellent (seeing: 0.7", a coherence time of 5$\,$ms, a wind speed of 6$\,$m\,s$^{-1}$ at 30$\,$m altitude).
The spectrograph divided the light into 3721 channels of $\Delta \lambda = 0.125\,$nm, ranging from $470\,$nm to $935\,$nm.
To process the data, we used the {\tt mpdaf.obj}-package\footnote{\href{https://mpdaf.readthedocs.io/en/latest/index.html}{https://mpdaf.readthedocs.io/en/latest/index.html}} \citep[][version~3.6]{Bacon2016}.
To generate an image with sufficient signal, we stacked the spectral channels.
Our analysis of the data revealed that the signals of interest, particularly those emanating from the large-scale environment of the FUor object, are predominantly found in the longer wavelength channels.
Therefore, in our stacking process, we selectively included the 1000 channels with the longest wavelengths (810--935$\,$nm).
We then binned these 1000 channels into 50 representative channels.
In each representative channel, we subtracted the average of columns and rows to remove the vertical and horizontal stripes seen on the detector.
A spectral analysis of the line emission will be carried out in a later study.

\subsection{VLT/SPHERE total intensity}
The $H$-band VLT/SPHERE image presented in \citet{Weber2023}, focused exclusively on the polarized component of the observed light. However, the same observation (098.C-0422, PI: L.~Cieza) also provides the total intensity (i.e., the sum of both beams on the detector).
To obtain the total intensity, we summed all the aligned object frames.
We produced the total intensity using the IRDAP reduction pipeline\footnote{\href{https://irdap.readthedocs.io/en/latest/}{https://irdap.readthedocs.io/en/latest/}} \citep[][version~1.3.5]{vanHolstein2020}.

\section{Results}\label{sec:results}
\subsection{ALMA continuum emission}
\begin{figure*}
    \centering
    \includegraphics[]{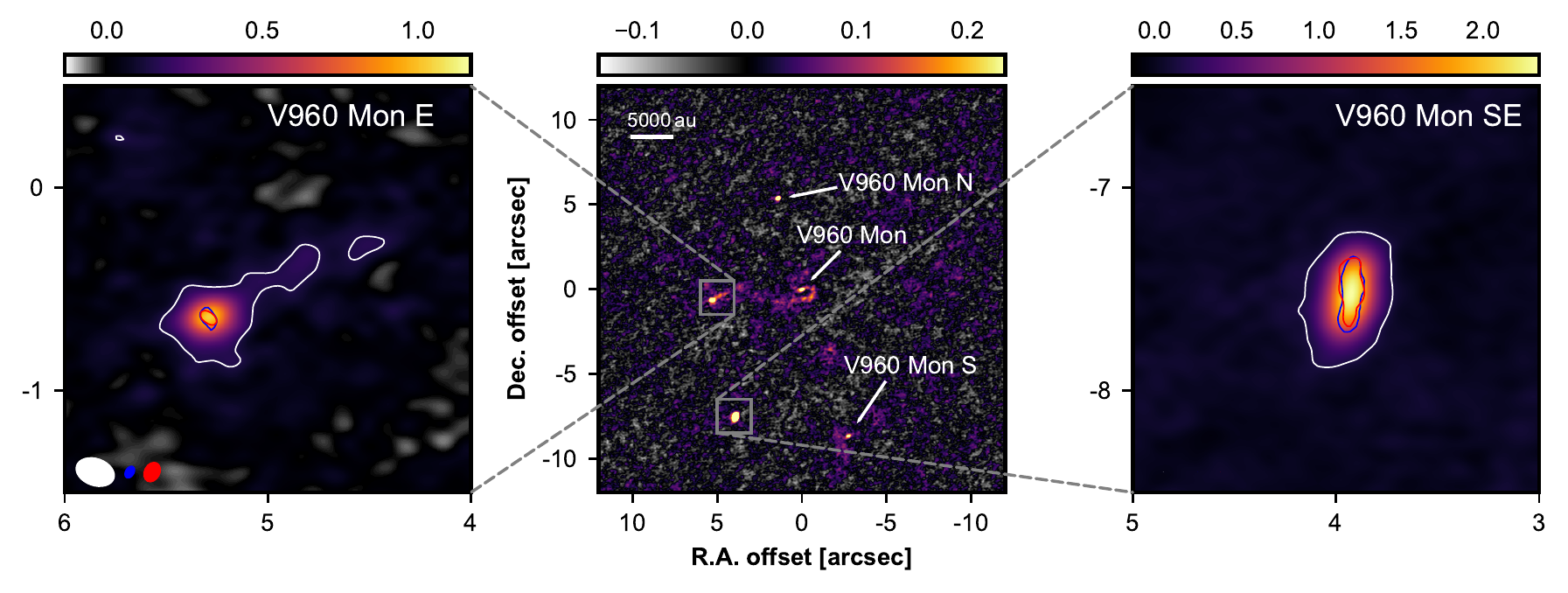}
    \caption{Continuum emission from the region around \vmon{}, in units of mJy$\,$beam$^{-1}$. {\it Center:} The large-scale view of the \bsix{} continuum around \vmon{}. Prominent sources of mm-emission are marked by arrows and labeled. {\it Left:} Detected emission named \vmonE{}. The image itself shows the \bsix{} data, white, red and blue contours are at 5$\,\sigma_{\rm rms}$ for \bsix{}, \bfour{} and \bthree{}, respectively. The ellipses in the bottom left corner show the restored beams for the respective color. {\it Right:} Same as left panel but for \vmonSE{}.}
    \label{fig:band6_cont_triple}
\end{figure*}

\vmon{} was observed by ALMA in \bsix{}, \bfour{} and \bthree{}.
In a previous study, \citet{Weber2023} reported a fragmenting spiral arm in the \bsix{} continuum.
Their analysis focused on unresolved clumps detected at roughly the 5\,$\sigma_{\rm rms}$ level.
We do not find any \bthree{} or \bfour{} counterparts to these clumps.
This non-detection is consistent with thermal emission from the clumps, which would yield fluxes of $\lesssim 2\,\sigma_{\rm rms}$ in \bfour{} and $\lesssim 1\,\sigma_{\rm rms}$ in \bthree{} for the rms-noise values listed in Table~\ref{tab:ALMA_obs}.

When inspecting the large-scale environment of the FUor, several bright sources appear in the FOV as shown in the \bsix{} image in the central panel of Fig.~\ref{fig:band6_cont_triple}.
Following \citet{Kospal2021}, we labeled these sources \vmon{} for the outbursting star at the center of the image, \vmonN{} for the northern object, \vmonE{} for the eastern object, \vmonSE{} for the southeastern object and V960\,Mon\,S for the southern object.
The sources, their projected position relative to \vmon{}, their continuum flux in the different bands, the corresponding estimated dust masses and spectral indices are listed in Table~\ref{tab:sources}.
To convert flux into mass, we assumed a dust temperature of 20\,K and an opacity of 
$\kappa_\nu = 1\,{\rm cm}^2\,{\rm g}^{-1}\,(\nu/100\,{\rm GHz})$.

\paragraph{Sources and Limits}
\vmonN{} has been detected several times \citep[see e.g.][]{Kospal2015}, as it is also present in the optical and NIR.
We detected it in all three analyzed continuum bands.

\vmonE{} was previously reported from the same ALMA \bsix{} data set in \citet{Kospal2021}, but was not confirmed in any other observations. 
Here, we found that it also shows significant emission in ALMA bands~3 and 4, as shown in the left panel of Fig.~\ref{fig:band6_cont_triple} where the contours show the emission at 5$\,\sigma_{\rm rms}$ for the different bands. 
The peak emission reaches $7\,\sigma_{\rm rms}$, $14\,\sigma_{\rm rms}$ and 
$40\,\sigma_{\rm rms}$ in bands~3, 4 and 6, corresponding to 0.161\,mJy\,beam$^{-1}$, 0.35\,mJy\,beam$^{-1}$ and 1.12\,mJy\,beam$^{-1}$.

The source \vmonSE{} is even more luminous.
It was first reported in \citet{Kospal2015}, where the authors showed that the compact emission does not appear for observational wavelengths $\lambda_{\rm obs} \leq 4.5\,$µm but already dominates the flux of the system at $\lambda_{\rm obs}=70\,$µm.
The right panel of Fig.~\ref{fig:band6_cont_triple} shows the 5$\,\sigma_{\rm rms}$ contours for the three different ALMA bands, centered on \vmonSE{}. 
In contrast to \vmonE{}, this source appears resolved in all three observations, at least along the North-South axis.
It clearly shows that the emission is elongated along this axis, ruling out emission from a single point source.
To estimate the structure's spatial extent, we fitted a two-dimensional Gaussian to the \bfour{} data.
The fit yielded full width at half maximum (FWHM) values for a major axis of $0\farcs30\pm 0\farcs02$ and a minor axis of $0\farcs085\pm 0\farcs01$.
If this continuum emission stems from a disk-like configuration, the corresponding disk radius would be $\sim280\,$au, with an inclination of $i_{\rm SE}\geq74^\circ\pm2^\circ$, and a position angle of $-0.3^\circ\pm1^\circ$.

Finally, we detected the unresolved source \vmonS{} in \bsix{} and \bfour{}, while \bthree{} does not yield a significant detection.

\paragraph{Spectral Index}
To investigate the nature of the detected emission, we calculated the spectral indices for the sources in the FOV, for different band combinations, following the formula:
\begin{equation}\label{equ:spectral_index}
    \alpha_{i/j} = \frac{\log{F_{i}/F_{j}}}{\log{\nu_i/\nu_j}}\,,
\end{equation}
where $F$ is the observed flux from an object in two distinct bands $i$ and $j$, and $\nu$ is the corresponding representative frequency.
We calculated the representative frequencies by averaging the respective values of the continuum spectral windows and find $\nu_3=93\,$GHz, $\nu_4=153\,$GHz and $\nu_6=224.5\,$GHz.
\begin{table*}[]
\centering
\begin{tabular}{lccccccccccc}
\tableline
Name & $R\,[^{\prime\prime}]$ & $P.A.\,[^{\circ}]$ & $F_3[{\rm mJy}]$ & $F_4[{\rm mJy}]$ & $F_6[{\rm mJy}]$ & $ M_3\,[M_\oplus]$  & $M_4\,[M_\oplus]$  & $M_6\,[M_\oplus]$ & $\alpha_{6/4}$ & $\alpha_{4/3}$ & $\alpha_{6/3}$\\
\tableline
V960 Mon & -- & -- & 0.045& 0.36 & 0.94& 40 & 77 & 71 &1.9$\pm0.2$& 2.3$\pm$0.5 & 2.1$\pm$0.3\\
N & 5.54 & 14.7 & 0.17 & 0.42 & 1.09 & 151 & 90 & 82 & 3.1$\pm$0.2 & 1.7$\pm$0.4& 2.3$\pm$0.2\\
E & 5.32 & 96.6 & 0.12 & 0.42 & 1.66 & 107 & 92 & 125 & 2.8$\pm$0.2 & 1.6$\pm$0.2 & 2.1$\pm$0.1\\
SE & 8.46 & 152.4 & 0.62 & 2.49 & 6.41 & 558 & 540 & 482 & 1.7$\pm$0.4 & 2.7$\pm$0.5 & 2.1$\pm$0.4\\
S & 9.05 & 197.7 & --& 0.20 & 0.82 & -- & 42 & 62 & 2.1$\pm$0.4 & -- &--\\
\tableline
\end{tabular}
\caption{List of mm-sources visible in ALMA data. The different columns list the distance to the primary, $R$, and the corresponding position angle $P.A.$, as well as the integrated fluxes measured for band$\,$3, band$\,$4 and band$\,$6, and the corresponding dust mass estimates, assuming optically thin emission at a temperature of 20$\,$K and a dust opacity of $\kappa_\nu = 1\,{\rm cm}^2\,{\rm g}^{-1}\,(\nu/100\,{\rm GHz})$. The spectral index $\alpha$ is defined in equ.~\ref{equ:spectral_index}. The estimated errors arise from the propagated rms-noise of each band listed in Table~\ref{tab:ALMA_obs}.}
\label{tab:sources}
\end{table*}
Table~\ref{tab:sources} lists the spectral indices calculated from different bands for the five most prominent flux sources in the \bsix{} FOV.
{The uncertainties given for the spectral indices in Table~\ref{tab:sources} stem from error propagation of the uncertainties in the flux measurement due to the rms-noise.}
We add a note of caution: while including \bsix{} in the spectral index calculation is advantageous—thanks to its higher S/N and greater frequency separation—it may also introduce a systematic error.
The \bsix{} observations were carried out approximately four years before the \bfour{} and \bthree{} data.
Since \vmon{} (and possibly other detected sources) appears variable, a variability on anual timescales would significantly compromise the spectral index analysis.
Band\,3 and \bfour{} data were taken on subsequent days. 

In source S, the only measurement of the spectral index, $\alpha_{\rm 6/4}$, is consistent with 2. 
In the remaining four sources, the values of $\alpha_{\rm 6/3}$ are also consistent with 2. 
These sources are likely optically thick at 225\,GHz with optical depth $\tau_{\rm 225 GHz}\gtrsim$5, similar to the bright Class-II disks in the Taurus-Auriga region \citep{Chung2024ApJS..273...29C}, which is expected by \citet{Delussu2024}.
We briefly discuss the better-detected sources -- V960\,Mon, N, E, and SE -- in the following.

\paragraph{\vmon{}, Source SE} The values of $\alpha_{\rm 6/4}$ are comparable to or slightly smaller than 2.
In the latter case, the maximum dust grain sizes, $a_{\rm max}$, may be close to $\sim$100 $\mu$m, causing the spectra to appear anomalously reddened due to dust scattering \citep{Liu2019ApJ...877L..22L}.
The values of $\alpha_{\rm 4/3}$ are larger than $\alpha_{\rm 6/4}$.
This may imply that the optical depths of the dominant (sub)millimeter emitters become $\sim$1--3 at the central frequency of our ALMA \bfour{} observations, $\sim$153\,GHz, although uncertainties remain depending on the assumed dust temperatures, $a_{\rm max}$, and dust opacity tables (cf. Section 5.2.1 of \citealt{Chung2024ApJS..273...29C}).

\paragraph{Source N, E} The values of $\alpha_{\rm 4/3}$ are likely somewhat smaller than 2, while the values of $\alpha_{\rm 6/4}$ are close to 2.5--3.
These are similar to what was observed in the archetypal FUor, FU\,Ori \citep{Liu2019ApJ...884...97L,Liu2021ApJ...923..270L}, which were also resolved in many Class 0/I YSOs \citep{Li2017ApJ...840...72L}.
There are two probable interpretations:
(1) Dust emission is very optically thick at $>$93 GHz frequencies, and the value of $a_{\rm max}$ in the dominant (sub)millimeter emitters is a few times 100\,µm (depending on the assumed dust composition and porosity).
In such cases, the dust albedo can increase around $\sim$100\,GHz and then decrease at higher frequencies (cf. Figure~3 in \citealt{Liu2019ApJ...877L..22L}).
This leads to anomalously reddened and bluened spectra at $\sim$100\,GHz and $\gtrsim$200\,GHz, respectively (for an example, see the SED model for DM\,Tau in \citealt{Liu2024A&A...685A..18L}). (2) Dust emission is very optically thick, and there is a temperature gradient in the line of sight and the warmer emitters are obscured by the colder ones.
This configuration may be realized for edge-on sources \citep{Galvan2018ApJ...868...39G}, or for sources that are dominantly heated by dissipative processes in the gaseous disks (e.g., viscous or shock heating, etc; \citealt{Liu2019ApJ...884...97L,Liu2021ApJ...923..270L,Zamponi2021MNRAS.508.2583Z}).
These two interpretations are not mutually exclusive.
To discern the values of $a_{\rm max}$, future spatially resolved observations at more frequency bands, or the (sub)millimeter linear polarization observations to probe signatures of dust self-scattering, are necessary. 
The observed, lower $\alpha_{\rm 4/3}$ in these two sources may also be partly explained by the contribution of optically thin free-free emission. 
Nevertheless, previous JVLA surveys toward the Corona Australis, Ophiuchus, and Orion molecular clouds showed that free-free emission in low-mass YSOs (without including proplyds around OB stars) is commonly not bright enough to contribute significantly at $>$90\,GHz frequencies \citep{Liu2014ApJ...780..155L,Coutens2019A&A...631A..58C,Tychoniec2021}. 

\subsection{Molecular line emission}\label{subsec:COemission}
\begin{table}[]
    \centering
    \begin{tabular}{lcccccc}
        \tableline
       Molecule & $J$ & $\nu$ & $v_{\rm res}$ & $E_{\rm up}\,$ & $\mu^2S$ & log$_{10}(A_{ij})$\\
       \tableline
       $^{12}$CO & 2--1 & 230.538 &0.04 &16.6 & 0.024& $-6.2$\\
       $^{13}$CO & 2--1 & 220.399 &0.08 &15.9 & 0.024& $-6.2$\\
       C$^{18}$O & 2--1 & 219.560 &0.08 &15.8 & 0.024& $-6.2$\\
       DCN       & 3--2 & 217.238 &0.67 &20.9 & 80.5 & $-3.3$\\
       DCO$^+$   & 3--2 & 216.113 &0.68 &20.7 & 142.0& $-2.6$\\
       N$_2$D$^+$& 3--2 & 231.322 &0.63 &22.2 & 312.1& $-3.15$\\
       SiO       & 5--4 & 217.105 &0.67 &31.2 & 48.0 & $-3.3$\\
       \tableline
    \end{tabular}
    \caption{Detected molecular transitions. Rest frequency is in GHz, velocity resolution in \kms, upper energy level in K and line strength ($\mu^2S$) in Debye$^2$. $A_{ij}$ is the Einstein coefficient for the respective transition.}
    \label{tab:molecules}
\end{table}
The spectral windows of the ALMA \bsix{} observation cover several molecular transition lines.
For line detection, we extracted spectra directly from the visibilities of each spectral window, identifying signals corresponding to rotational transitions of $^{12}$CO, $^{13}$CO, \ceo{}, \sio{}, \dcop{}, \ntdp{} and DCN. 
This approach is effective for detecting the brightest emission lines, as visibilities inherently integrate spatial information including noise.
To achieve more robust detections and isolate emission from individual sources, we subsequently performed the analysis on the reconstructed data cubes for each line.
These molecules, the detected rotational transitions, their reference frequency, the velocity resolution of the observation in the respective spectral window, and the upper energy level of the transitions are listed in Table~\ref{tab:molecules}.

\subsubsection{CO isotopologues}
We measured the rms-noise per respective channel width of $\sigma_{\rm rms}(^{12}{\rm CO}) = 9.3\,{\rm mJy}\,{\rm beam}^{-1}$, $\sigma_{\rm rms}(^{13}{\rm CO}) = 7.0\,{\rm mJy}\,{\rm beam}^{-1}$ and $\sigma_{\rm rms}({\rm C}^{18}{\rm O}) = 6.6\,{\rm mJy}\,{\rm beam}^{-1}$.
We collapsed the CO data cubes and calculated the moment-0 (integrated intensity) and moment-1 (intensity‐weighted mean velocity) maps using the {\tt bettermoments} package \citep{bettermoments}.
For both moments, we applied sigma clipping below 2$\,\sigma_{\rm rms}$ and smoothed the threshold mask over one beam size to avoid steep edges in the mask.
\begin{figure*}
    \centering
    \includegraphics[]{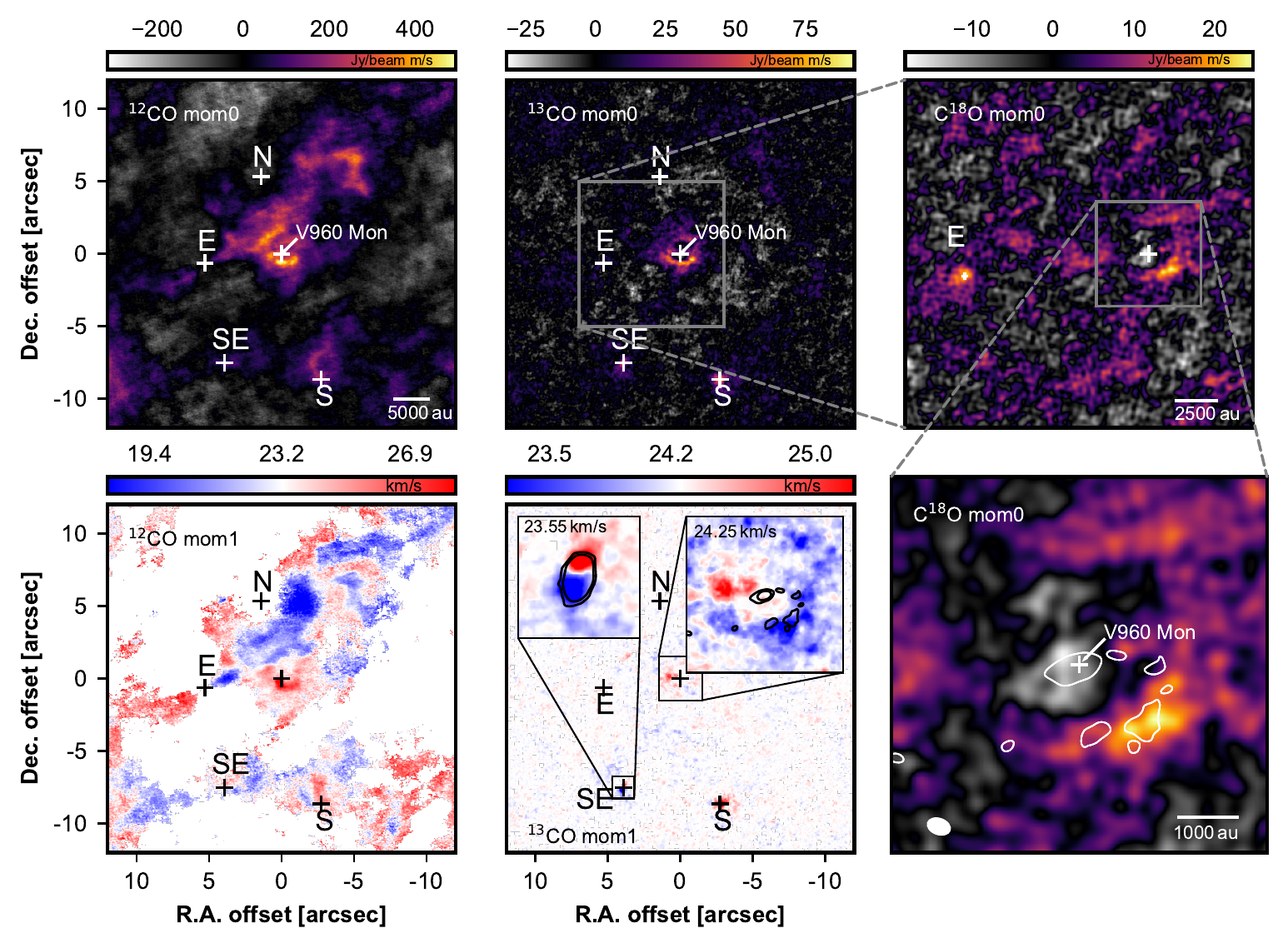}
    \caption{CO(J=2$-$1) line emission in Jy\,beam$^{-1}$~m\,s$^{-1}$ for the three different isotopes $^{12}$CO (left), $^{13}$CO (center) and \ceo{} (right) at different spatial scales. The top left and center show the moment-0 maps for $^{12}$CO and $^{13}$CO, respectively, with the bottom panels showing the corresponding moment-1 map.
    The crosses indicate the locations of the \bsix{} continuum emission.
    The bottom center panel includes insets with black contours at $5\sigma$ and $10\sigma$ of the \bsix{} continuum emission.
    The color scale of the insets ranges from the displayed velocity to $\pm1$\kms{}.
    The right column shows the \ceo{} moment-0 map at different scales, with the lower panel focusing on the area around \vmon{}.
    The white contours show the \bsix{} continuum at $5\sigma$ {and the white ellipse in the bottom left corner marks the clean beam of the observation}.}
    \label{fig:12CO}
\end{figure*}
\begin{table}[]
    \centering
    \begin{tabular}{lcc}
        \tableline
       Object & $v_{\rm los}\,$[\kms] & molecule \\
       \tableline
       \vmon{}  & 24.25 & $^{12}$CO\\
       \vmonE{}  & 23.19 & $^{13}$CO\\
       \vmonSE{}  & 23.55 & $^{12}$CO\\
       \vmonS{}  & 26.38 & $^{12}$CO\\
       \tableline
    \end{tabular}
    \caption{Line of sight velocities for different objects and the employed molecule (see Appendix~\ref{appendix:los}).}
    \label{tab:vlos}
\end{table}
Fig.~\ref{fig:12CO} shows the resulting moment-0 maps of the CO isotopes at different spatial scales and moment-1 maps for $^{12}$CO and $^{13}$CO.
From these data, we inferred the line of sight velocities listed in Table~\ref{tab:vlos} and described in Appendix~\ref{appendix:los}.

\paragraph{$^{\it 12}$CO}

The {left} column of Fig.\ref{fig:12CO} presents the $^{12}$CO isotopologue, where significant foreground emission obscures the region surrounding \vmon{}.
We estimated the rms-noise of the moment-0 map to be $\sigma_{\rm rms}(^{12}{\rm CO})=20\,{\rm Jy}\,{\rm beam}^{-1}~{\rm m}{\rm s}^{-1}$.
Examination of the $^{12}$CO kinematics in the central left bottom panel reveals blue- and red-shifted lobes spectrally centered around a velocity of approximately $22.3$\kms{}, and spatially centered on the mm-continuum source designated as \vmonE{}.
This pattern suggests a projected outflow velocity of about $3.5$\kms{}, originating from the unclassified object \vmonE{}.
We further discuss this feature in Sec.~\ref{subsec:species} and Sec.~\ref{sec:discussion}.
\begin{figure}
    \centering
    \includegraphics[width=\columnwidth]{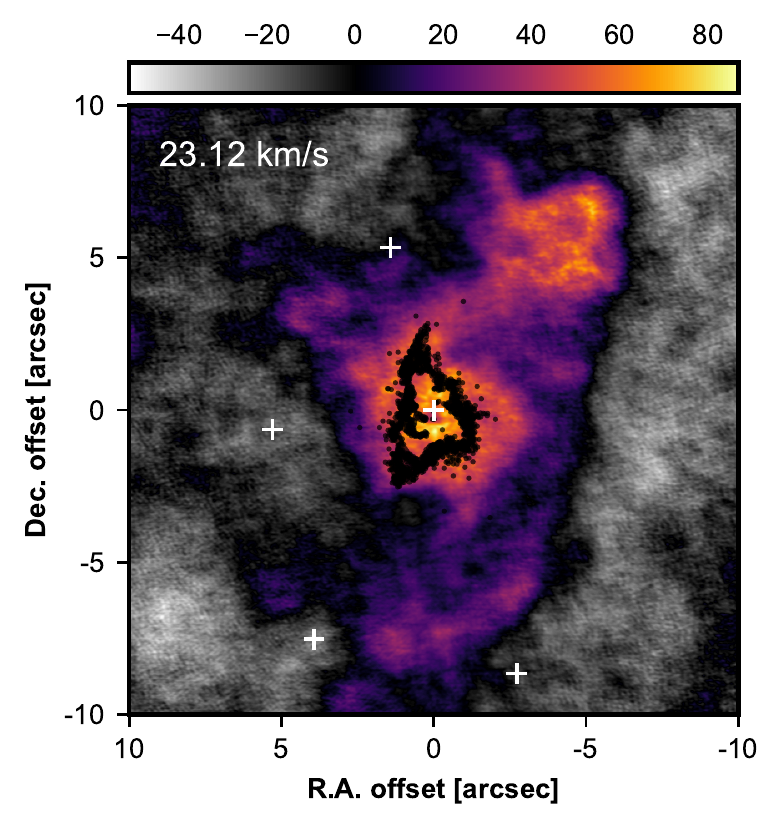}
    \caption{$^{12}$CO emission at 23.12\kms in Jy\,beam$^{-1}$~m\,s$^{-1}$. The black contours delineate the spiral structure observed in NIR polarized light \citep{Weber2023}. The white plus signs indicate the sources of mm-emission in the FOV.}
    \label{fig:12COchannel140}
\end{figure}

When integrating over the entire spectral axis, multiple emission components blend along the line of sight.
Fig.~\ref{fig:12COchannel140} shows the emission of $^{12}$CO recorded at 23.12\kms{} (we stacked channels from 22.92\kms to 23.32\kms to increase the SNR).
At these velocities, the signal shows gas around the primary, while also including what seems to be interaction with slightly blue-shifted gas reservoirs to the northwest and south.
For comparison, the black contours show the polarized light of the SPHERE $H$-band image \citep{Weber2023}, which delineate a spiral structure possibly related to the large-scale interaction.

\paragraph{$^{\it 13}$CO}
The central panels of Fig.~\ref{fig:12CO} show the moments of the optically thinner $^{13}$CO emission.
We estimated the noise of the moment-0 map to be $\sigma_{\rm rms}(^{13}{\rm CO})=7\,{\rm Jy}\,{\rm beam}^{-1}~{\rm m}{\rm s}^{-1}$.
Within the field of view, we identified three principal regions of $^{13}$CO emission: in the vicinity of the primary source \vmon{}, near \vmonSE{}, and around \vmonS{}.
Notably, all three regions coincide with \bsix{} continuum emission at levels above $10\,\sigma_{\rm rms}$.
The bottom {central} panel of Fig.~\ref{fig:12CO} shows the moment-1 map of the $^{13}$CO line, where a blue/red-pattern centered on the continuum emission of \vmonSE{} becomes apparent, suggesting rotation around a massive compact source.
Around \vmon{}, we found red-shifted contributions towards the northeast, while the clumps in the southwest are co-located with blue-shifted line emission.

\paragraph{C$^{\it 18}$O}
The right column of Fig.~\ref{fig:12CO} displays the moment-0 map of detected \ceo{} emission around \vmon{}.
The measured noise of the map is $\sigma_{\rm rms}({\rm C}^{18}{\rm O})=3\,{\rm Jy}\,{\rm beam}^{-1}~{\rm m}{\rm s}^{-1}$.
The top right panel shows the region 
including \vmon{} and the E source. 
Emission from the \vmonE{} is detected with a maximum of $23.2\,{\rm Jy}\,{\rm beam}^{-1}~{\rm m}{\rm s}^{-1}$.
The bottom right panel focuses on the area closer to \vmon{}, coinciding with the fragmented continuum spiral presented in \citet{Weber2023}.
The \ceo{} emission at the clumps' location reaches a maximum value of $25.0\,{\rm Jy}\,{\rm beam}^{-1}~{\rm m}{\rm s}^{-1}$.
The detection of \ceo{} suggests a substantial gas content associated with the clumps, indicating that the fragmentation is not confined solely to the dust phase.

\subsubsection{Further species}\label{subsec:species}
\begin{figure*}
    \centering
    \includegraphics[width=\textwidth]{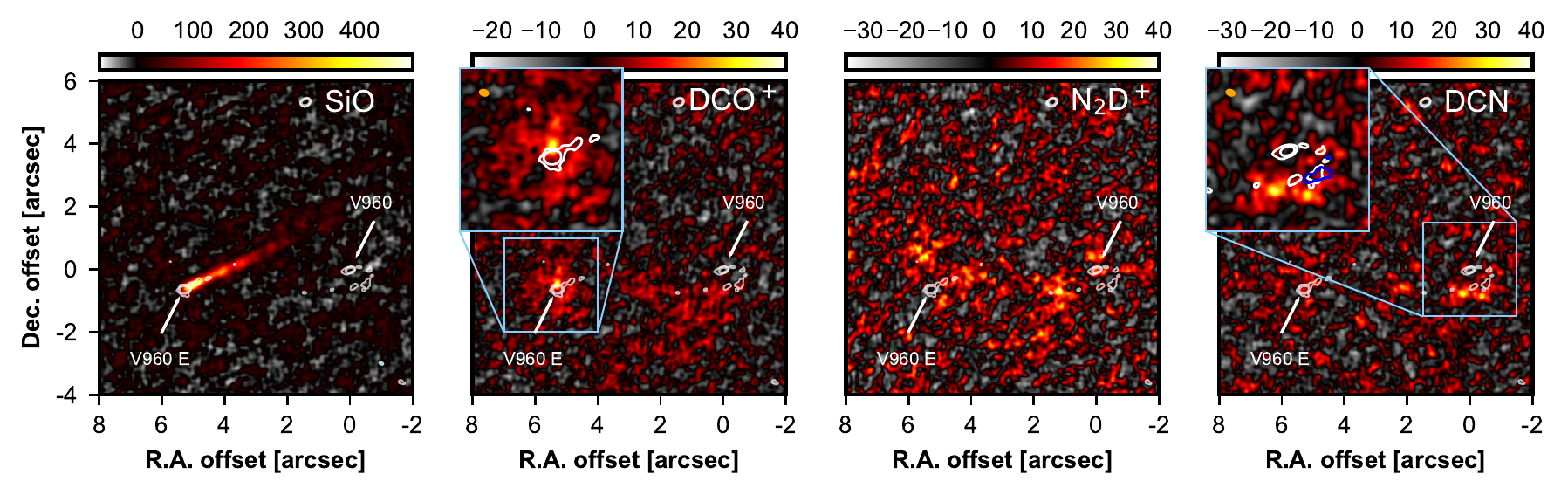}
    \caption{Moment-0 maps of molecular tracers in the system of \vmon{}. The units are mJy$\,$beam$^{-1}$~${\rm m}{\rm s}^{-1}$. The panels show the area between \vmonE{} and \vmon{} (there is no relevant detection outside this area). In all panels, the white contours trace the \bsix{} continuum flux at 5 and 10\,$\sigma_{\rm rms}$. In the inset of the right panel, the blue contours mark the \ceo{} at 5\,$\sigma_{\rm rms}$.{In both the insets of center-left and right panels, the orange ellipse in the top left corner shows the clean beam of the observation.}}
    \label{fig:molecules}
\end{figure*}
In Fig.~\ref{fig:molecules} we show the tracers, SiO, \dcop{}, \ntdp{} and DCN, that we detected in the spectral windows that were dedicated to the continuum emission.
To produce the moment-0 map we used {\tt bettermoments} without any sigma-clipping or smoothing.
The following paragraphs discuss the distribution and significance of each tracer.
The deuterated species are thought to be a product of deuterium fractionation that distributes deuterium from HD into other species \citep{Nomura2023}.
The deuterium fraction is expected to be significantly enhanced in cold areas ($T\lesssim30\,$K), the detection of deuterium can therefore inform about the local temperature.
Nevertheless, the different deuterated species are detected at very different locations in the image, necessitating investigation of each molecule individually.
As seen from Table~\ref{tab:molecules}, \dcop{}, \ntdp{} and DCN all have similar upper energy states ($E_{\rm up}\simeq 20\,$K).
This implies that differences in the distribution of their emission arise primarily due to different local abundances and line strengths, rather than excitation temperatures.

\begin{figure}
    \centering
    \includegraphics[width=\columnwidth]{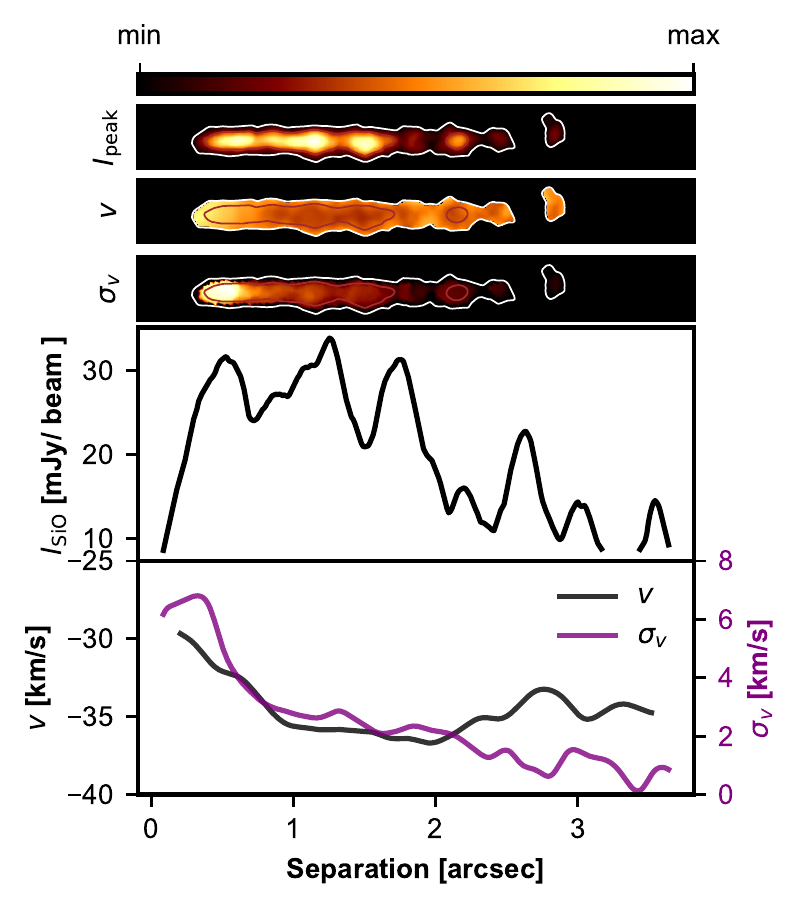}
    \caption{Analysis of the SiO jet rotated by 25.5$^\circ$. The top three panels show the distributions of the flux at peak emission, the moment-1 and moment-2, respectively. {They share the same colorbar of linear scale with minimum to maximum ranges of [0 to 34]\,mJy$\,$beam$^{-1}$, [$-45$ to $-26$]\kms{} and [0 to 7]\kms, respectively.} In all three panels, the contours trace $I_{\rm peak}$ at 10 and 20$\,\sigma_{\rm rms}$. For a quantitive analysis, we plot the maxium intensity, the minimum velocity relative to the systematic velocity of \vmonE{} and the maximum velocity dispersion, $\sigma_v$, in the lower two panels.}
    \label{fig:SiOjet}
\end{figure}

\paragraph{\sio{}}
The left panel of Fig.~\ref{fig:molecules} displays the moment-0 map of Silicon monoxide (\sio{}) with an rms-noise of $\sigma_{\rm rms}({\rm SiO})\approx8.8\,{\rm Jy}\,{\rm beam}^{-1}~{\rm m}{\rm s}^{-1}$.
The image shows a strong and collimated emission feature attached to \vmonE{}.
The peak emission of the \sio{} feature is shifted by $-14.8$\kms{} with respect to the restframe, so about $-$37.1\kms{} relative to the systematic velocity of \vmonE{} ($v_{\rm E}=$22.3\kms).
Such collimated \sio{} emission is typically linked to a high velocity jet shocking with the surrounding medium.
This is characteristic for actively accreting class-0 stars \citep[e.g.][]{Codella2007}. 
While the \sio{} feature is spatially very confined, it is widely spread over different velocity channels, with a typical FWHM line width of about 10\kms{} (from $-30$\kms{} to $-40$\kms{} with respect to \vmonE{}).
These broad emission lines are typical of shocked regions \citep[e.g.][]{Bachiller1997}.
For Fig.~\ref{fig:SiOjet}, we rotated the jet by $-25.5^\circ$ to align it horizontally.
In the three upper panels of Fig.~\ref{fig:SiOjet}, we show the peak intensity, moment-1 and moment-2 (intensity weighted velocity dispersion, $\sigma_v$), overlayed with the peak intensity contours at 10, 20 and 40$\,\sigma_{\rm rms}$.
The white contours display that the width of the jet is roughly constant with separation.
We fit a Gaussian function to the jet width and retrieve a FWHM of $0\farcs21$, similar to the \bsix{} beam size.
For the two lower panels of Fig.~\ref{fig:SiOjet}, we extract the maximum values (minimum, for moment-1) along the vertical direction in the top panels.
To reduce the spiky structure of the velocity profiles, we smoothed the distribution with a Gaussian kernel.
The intensity profile highlights that the emission is distributed in several knots along the jet. 
Similar behavior has been detected in SiO jets for several other class-0 objects \citep[e.g.][]{Codella2007,Podio2016,Wang2019}.
Here, we measure a periodic reoccurrence of the knots with a distance of about 0$\farcs$55 between consecutive maxima and a clear detection of at least seven peaks.
In the bottom panel of Fig.~\ref{fig:SiOjet} we show that the relative line of sight velocity to \vmonE{} becomes more blue-shifted with distance, ranging from about $-30$\kms{} at the base of the jet, to about $-40$\kms{} at distances larger than $1''$.
However, this effect may not represent a genuine acceleration but could instead reflect jet rotation within $1''$.
By contrast, the velocity dispersion is greatest near to \vmonE{} ($\sigma_v \sim7$\kms{}) and declines with distance.
To estimate the time interval between jet expulsions from the angular separations measured in the image plane, we assume that the jet velocity in the plane of the sky is comparable to its line of sight velocity ($\sim 40$\kms) and that all the knots have this same constant velocity.
From this, we find that the knots correspond to expulsion events roughly every 150$\,$ years, consistent with the periodicity found for molecular jets in other systems \citep[see][for a review]{Lee2020}.
This variability implies that \vmonE{} is episodically accreting.

\paragraph{DCO$^{+}$}
We detected the deuterated formylium ion (DCO$^{+}$) in the vicinity of \vmonE{}.
The moment-0 map presented in the center-left panel of Fig.~\ref{fig:molecules} has a rms-noise of $\sigma_{\rm rms}({\rm DCO}^{+})\approx3.0\,{\rm Jy}\,{\rm beam}^{-1}~{\rm m}{\rm s}^{-1}$.
Notably, the DCO$^{+}$ emission is spatially offset from the continuum source by $0\farcs29$, corresponding to a projected distance of approximately 644$\,$au.
The presence of DCO$^{+}$ requires some amount of gas-phase CO and low temperatures.
The asymmetry of the DCO$^{+}$ emission near \vmonE{} might result from low SNR or indicate an intrinsically non-uniform environment, possibly shaped by outflows from \vmonE{} itself and by the complex gravitational interactions with the surrounding system.
We further detect some \dcop{} emission from around \vmonS{} which is shown in Appendix~\ref{appendix:spectrum}.

\paragraph{\ntdp}
The center right panel of Fig.~\ref{fig:molecules} shows the moment-0 map of \ntdp{} with an rms-noise of $\sigma_{\rm rms}({\rm SiO})\approx5.4\,{\rm Jy}\,{\rm beam}^{-1}~{\rm m}{\rm s}^{-1}$.
The ion \ntdp{} is destroyed by reactions with CO, inhibiting its detection in CO-rich regions \citep[e.g.][]{Caselli2002}.
Hence, \ntdp{} emission is strongly anti-correlated with CO.
Such conditions typically arise in cold environments where CO freeze-out onto dust grains prevails, supporting both the presence of \ntdp{} and its deuterium enrichment.
In contrast to the more spatially localized \dcop{} emission, the detected \ntdp{} distribution is relatively diffuse, appearing as sparse emission to the northeast of \vmonE{} and in the region between \vmonE{} and \vmon{}.
The emission is only marginally above 3\,$\sigma_{\rm rms}$ in four channels and primarily appears in the channel centered at 23.72\kms{} (see Appendix~\ref{appendix:spectrum}).  
Notably, the detected \ntdp{} signal locally overlaps with faint continuum \bsix{} emission, suggesting the presence of a potential bridging structure. Such filamentary connections have been discussed in the context of multiple stellar-formation scenarios \citep[e.g.,][]{Kuffmeier2019,Gieser2024} and in cases involving stellar flybys \citep[e.g.,][]{Cuello2020,Menard2020}. 

\paragraph{DCN}
The right panel of Fig.~\ref{fig:molecules} shows the moment-0 map of DCN with an rms-noise of $\sigma_{\rm rms}({\rm DCN})\approx4.1\,{\rm Jy}\,{\rm beam}^{-1}~{\rm m}{\rm s}^{-1}$.
The DCN emission is co-located with portions of the fragmented continuum spiral arm around \vmon{}.
The detection is based on four adjacent spectral channels between 23.03--25.05\kms{}, with the maximum emission at 23.70\kms{} (see Appendix~\ref{appendix:spectrum}).
In the right panel of Fig.~\ref{fig:molecules}, we overlay the moment-0 contours of the \ceo{} transition on the DCN emission.
This comparison highlights a clear anti-correlation: DCN is more prominent in the outer regions of the spiral, whereas it is depleted toward the inner clumps, where \ceo is strongest.
The presence of DCN in the outer clumps implies a substantial gas reservoir at these locations, and the \ceo{} detection farther inward suggests that gas extends along the entire fragmented continuum spiral.
Furthermore, the detection of DCN—alongside the non-detections of DCO$^{+}$ and N$_2$D$^{+}$—implies that this region has recently evolved from a very cold environment to one that is slightly warmer.
The ionized deuterated species such as DCO$^{+}$ and N$_2$D$^{+}$ are more easily depleted than DCN once the temperature rises \citep[e.g.][]{Rodgers1996,Aikawa2005,Liu2015}.
Therefore, the observed DCN emission, combined with the lack of DCO$^{+}$ and N$_2$D$^{+}$, supports a scenario in which the local chemistry is transitioning from a cold to a mildly warmer state.
This is consistent with the recent FUor event, which is expected to heat the surrounding region through irradiation.
However, a temperature increase could also be connected to contraction of the related clumps.\\ 
\quad 

As stated in \citet{Tychoniec2021}, the protostellar envelope is well-traced by \ceo{}, \dcop{} and \ntdp{}, all of which are detected around \vmonE{}.
The different locations and distributions at which we detect the signal indicate warmer and colder regions, characterized by the freeze-out of CO. 
Additionally, it is noteworthy that we do not detect any significant $^{13}$CO or \ceo{} emission above the noise level centered on the outbursting star \vmon{}.
This might be due to large-scale contamination in the image that is filtering out the emission from the primary.
The deuterated lines originate in smaller regions, making them less affected by this.
The northern companion, \vmonN{}, shows no trace of any molecular tracer, nor does it show any connections to other objects in the field of view.
This raises the question if \vmonN{} is truly associated with \vmon{}, or whether it is a fore- or background star.

\subsection{MUSE imaging}
\begin{figure*}
        \centering
    \includegraphics[width=\textwidth]{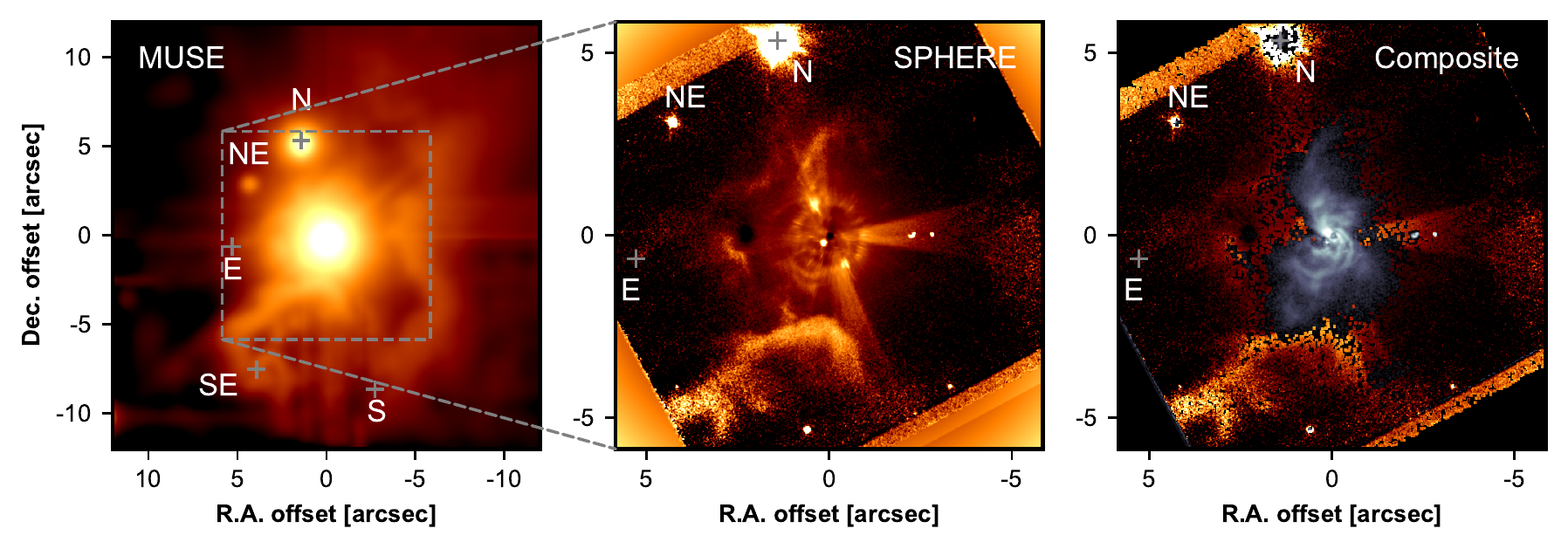}
    \caption{Left panel: VLT/MUSE image of \vmon{} and the region around it. Center panel: VLT/SPHERE $H$-band total intensity scaled with $r^2$ to highlight faint signal at large scales. Note the reduced field of view due to the smaller detector of SPHERE. Right panel: SPHERE total intensity image overlaid with SPHERE polarized intensity image \citep[grayscale, see][]{Weber2023}.}
    \label{fig:I_tot}
\end{figure*}
The left panel of Fig.~\ref{fig:I_tot} displays the image obtained by stacking the MUSE channels as described in Sec.~\ref{subsec:data_MUSE}.
Across the entire MUSE detector field of view ($65\farcs6\times 67\farcs6$), we detected 24 point sources.
Within the area covered by the ALMA \bsix{} primary beam (${\rm FWHM}\sim26''$), only \vmon{} and \vmonN{} among the \bsix{} continuum sources have a clear MUSE counterpart.
In addition, we find a faint point source in the northeast of \vmon{}, which we label \vmonNE{}.

In the vicinity of the primary object (at separations $\lesssim 3''$), \vmon{}’s luminosity dominates at these shorter wavelengths.
However, the MUSE image reveals intriguing large-scale structures.
Notably, the source identified as \vmonSE{} appears as the nucleus of a diffuse clump or cloudlet, intriguingly connected to the primary system via a structure of intertwined gas.
Remarkably, this represents the first detection of material associated with \vmonSE{} at wavelengths below 4.5$\,$µm, seen in reflection rather than in emission.
Previously, \vmonSE{} was reported as undetected at these shorter wavelengths \citep{Kospal2015}.
The observed decrease in brightness with increasing distance from \vmon{} suggests that the structure is primarily illuminated by the central FUor object, rather than being self-luminous at these wavelengths. 

Meanwhile, the enigmatic object to the east, \vmonE{}, shows no visible counterpart in the MUSE image.
One plausible explanation is that material closer to the primary star casts a shadow over \vmonE{}, preventing direct illumination from the FUor object due to the observed geometry.

To the west of \vmon{}, the image reveals what appears to be the illuminated interior of a surrounding envelope.
This observation corroborates the subtle emission noted in the $^{13}$CO moment-0 map, as shown in the central column of Fig.~\ref{fig:12CO}.
While the molecular lines trace emission, the MUSE counterpart likely shows light reflected from mircon-sized grains entrailed in the structure.
The observed structure is present in a large wavelength range, excluding the possibility of being associated with distinct molecular line emission in the optical and NIR.

Finally, the MUSE image in Fig.~\ref{fig:I_tot} underscores the value of stacked MUSE data in probing the interactions between young stars and their large-scale environments.

\subsection{SPHERE total intensity}\label{subsec:SPHERE_tot}
Polarized intensity ($I_{\rm pol}$) is a measure for imaging circumstellar environments because initially unpolarized light becomes polarized after scattering off dust grains.
Consequently, observations of polarized light trace the surfaces of dusty environments. 
However, this technique has several limitations.
First, by design, polarized light observations do not capture light directly emitted from companions (which is generally unpolarized), nor do they capture light scattered from unresolved dust structures around the light source, such as that expected from a circumstellar or circumplanetary disk. However, if this disk is sufficiently inclined or very asymmetric, the observed light from the central unresolved beam could be polarized.
Therefore, this method is not suitable for direct imaging of companions and can only detect them if their emission becomes net-polarized within the telescope's point spread function along the light path \citep[see e.g.,][for a discussion]{vanHolstein2021}.
Furthermore, the assumption that scattered light contributes to the polarized intensity relies on single scattering of photons.
If the majority of stellar photons in a certain region undergo multiple scatterings before escaping the system toward the observer, the emergent light becomes unpolarized due to the summation over randomly oriented polarization states.

Consequently, we analyzed the total intensity in which photons from companions and multiple scattering events still contribute, even if at a contrast that is difficult to detect.
The central panel of Fig.~\ref{fig:I_tot} displays the VLT/SPHERE $H$-band total intensity image, where the flux is scaled by the squared distance from the center to emphasize large-scale features.
A bright elongated structure appears to the south of \vmon{}.
Additionally, in the right panel of Fig.~\ref{fig:I_tot}, we show the polarized light $H$-band image \citep[presented in][]{Weber2023} in black and white.
This comparison shows that the elongated total intensity feature disappears approximately at the distances to where the spiral structures observed in scattered light extend.
Notably, we do not detect any polarization from the $H$-band counterpart, even after applying the same $r^2$-scaling.
Here, we discuss the reason for why this might be.
Even if the assumption of single-scattering holds true, the efficiency of polarization in a scattering event still depends on the scattering angle $\theta_{\rm scat}$.
In particular, polarization is predicted to be suppressed for backward scattering ($\theta_{\rm scat} > 90^\circ$) \citep[e.g.,][]{Min2005}.
Observations of multiple disks confirm a decrease in polarization efficiency at large scattering angles \citep[e.g.,][]{Ginski2023}.

Consequently, the lack of polarized light from the elongated feature in \vmon{} indicates a low polarization efficiency, likely due to backward scattering of the primary’s light.
This, in turn, implies that the elongated feature is situated behind the star from our viewpoint.
In fact, the feature disappears closer to the star, where the spiral structure is visible in scattered light, a result consistent with the spirals' optically thick region blocking the line of sight. 

Finally, we also note that although \vmonE{} lies within the SPHERE field of view, as in the MUSE image, it is detected in neither the total intensity nor the polarized intensity at $H$-band.
We conclude that \vmonE{} is deeply embedded, causing its flux to be heavily attenuated at short wavelengths. 

\section{Discussion}\label{sec:discussion}
The different observations of \vmon{} reveal a dynamic system around an FUor star, including several companions, spiral structures, clumps, and a connection to the large-scale environment.
The concurrence of these effects paints a complicated picture, where it remains unclear which process causes which effect.
This section considers several scenarios that connect the presented observations with concepts of star- and planet formation.
In particular, we will discuss the possibilites of large-scale infall perturbing the circumstellar environment (Sec.~\ref{subsec:infall}), a possible stellar flyby (Sec.~\ref{subsec:flyby}) and the ejection of compact objects from around \vmon{} by a stellar intruder to the circumstellar environment (Sec.~\ref{subsec:ejection}).

\subsection{Fragmentation and outburst induced by infall?}\label{subsec:infall}
Assuming the Gaia-measured distance to \vmon{} is accurate, the VLT/SPHERE image in scattered light \citep[][]{Weber2023} shows material present around \vmon{} up to a separation of $\gtrsim2000\,$au.
How such a vast environment can be sustained is an intriguing question.
As suggested by the detection of the elongated feature (see Fig.~\ref{fig:I_tot}, Sec.~\ref{subsec:SPHERE_tot}), one possibility is that \vmon{} is currently being fed by asymmetric infall from the neighboring SE source, separated by about 8\farcs4.
While infalling streamers onto circumstellar environments have been increasingly reported in recent years \citep[e.g.,][]{Ginski2021,Garufi2022,Cacciapuoti2024}, this could mark an interesting case where the associated reservoir of the streamer's origin is observed.
From the optical and NIR evidence alone, it is not discernible whether the feature connecting \vmonSE{} and \vmon{} represents infall onto the FUor.
A similar feature can also be observed during a flyby in which the gravitational interaction between the two objects creates an over-dense bridge \citep[e.g.][]{Cuello2020},
or from colliding flows in dense star formation regions \citep[e.g.][]{Kuffmeier2019}.

Guided by the detection in the optical and NIR, we investigated the ALMA \bsix{} cube for molecular counterparts.
From a channel-by-channel analysis, we found faint $^{13}$CO signal between 21.17\kms{} and 22.36\kms{} co-located with the MUSE elongated feature.
This implies that the line of sight velocity in the connecting gas is blue-shifted by $2-3$\kms{} with respect to \vmon{}.
We combine the channels in this velocity range of the cube using {\tt bettermoments}, masking out signal below 2$\,\sigma_{\rm rms}$ in each of the 14 channels.
The resulting moment-0 map limited to this velocity range has an rms-noise of $\sim3$\,Jy\,beam$^{-1}$~m\,s$^{-1}$.
\begin{figure}
    \centering
    \includegraphics[width=\columnwidth]{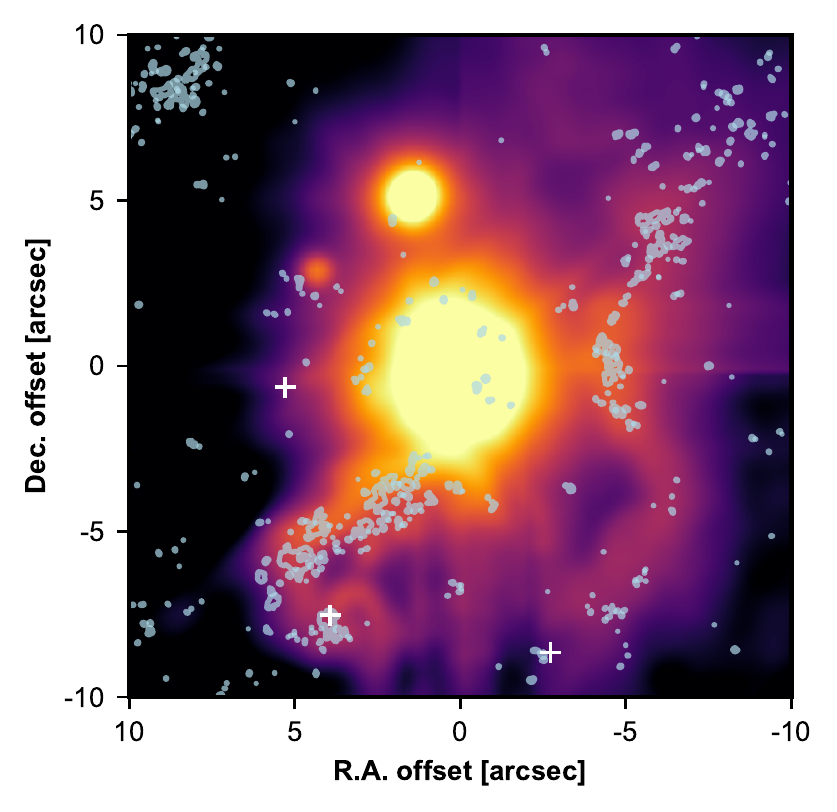}
    \caption{VLT/MUSE image (same as left panel of Fig.~\ref{fig:I_tot}). The blue contours show the $^{13}$CO-emission integrated between $21.17-22.36$\kms (blue-shifted by $\sim 2-3$\kms{} towards \vmon{}), where emission from the elongated feature can partly be separated from the contamination of the cloud at 2\,$\sigma_{\rm rms}$.}
    \label{fig:13filament}
\end{figure}
Fig.~\ref{fig:13filament} shows the MUSE image superimposed by the emission contours from these combined $^{13}$CO channels at 6\,Jy\,beam$^{-1}$~m\,s$^{-1}$ (corresponding to $\sim 2\sigma_{\rm rms}$).
In addition to the emission indicating the connection between \vmon{} and \vmonSE{}, there is faint $^{13}$CO emission at these velocities west of the primary, which could trace the inner wall of the surrounding envelope.

Having concluded in Sec.~\ref{subsec:SPHERE_tot} that the elongated structure is likely back-grounded with respect to \vmon{}, the blue-shifted velocity of the gas in the feature indicates that it is moving towards the outbursting star \vmon{}.
Hence, the combined analysis of optical, NIR and $^{13}$CO data gives evidence that the observed feature indeed represents infall onto the primary.

\citet{Kuffmeier2018} demonstrated that in a circumstellar environment the infall of material can locally reduce the Toomre parameter, leading to a localized GI and the formation of spiral structures in the disk around the star.
However, the simulation results presented there focused on an area much closer to the star.
The very extended structure around \vmon{} raises the question of whether the spirals are part of a gravitationally unstable disk or if they arise from infalling material that becomes twisted as it moves inward.
Testing a connection between the spirals and disk instability requires measuring underlying Keplerian rotation, evidence that enhanced $^{13}$CO observations could provide.
Moreover, although \citet{Kuffmeier2018} showed that localized GI induced by infall can generate a spiral structure, no fragmentation was observed in their simulations.
However, \citet{Longarini2023} suggest that in certain scenarios, fragmentation occurs only in the dust phase, a finding consistent with the ALMA continuum observation and not at odds with \citet{Kuffmeier2018}, where dust was not implemented.
If the spiral structure arises from non-Keplerian motion, it remains an open question under what conditions the infalling material might fragment.

GI in the outer parts of a protoplanetary disk can generate significant angular momentum transport.
This may lead to an increased gas flow towards the inner disk region where the switching on and off of the magnetorotational instability may trigger outbursts \citep{Armitage2001,Zhu2009,Stamatellos2011}.
If the disk gas is replenished via infall, outbursts could also be initiated due to GI-enhanced accretion \citep{Kuffmeier2018} or due to clump formation by fragmentation and subsequent accretion \citep{Vorobyov2015}.

\subsection{A possible stellar flyby?}\label{subsec:flyby}
\begin{figure}
    \centering
    \includegraphics[]{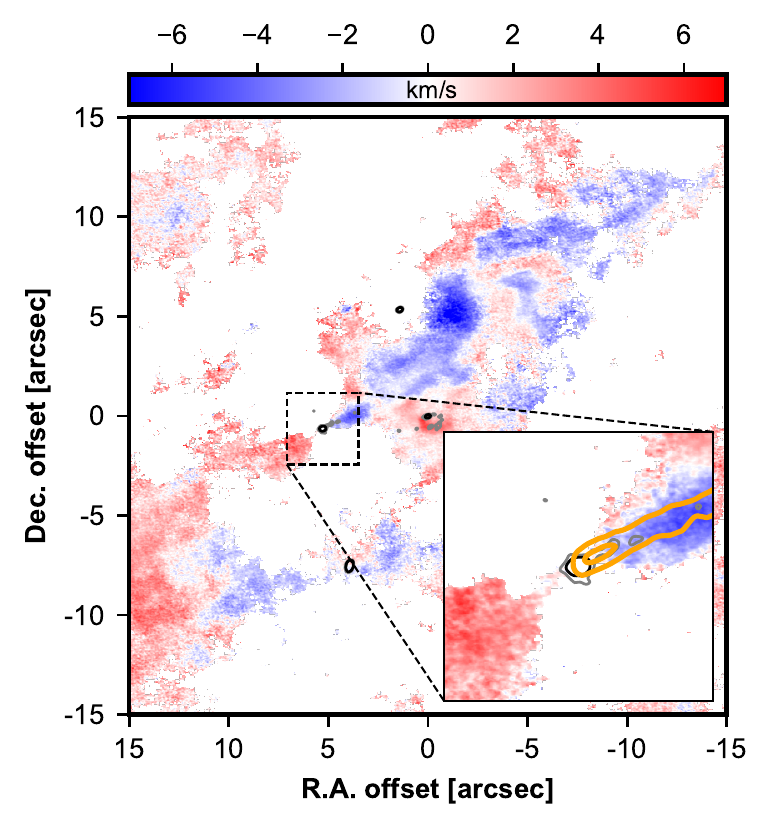}
    \caption{$^{12}$CO moment-1 {map, limited to where the $^{12}$CO moment-0 is positive}. {The velocity map is centered on the systematic velocity inferred for \vmonE{}, $v_{\rm E}=23.19$\kms.} The zoom-in is centered on the enigmatic object \vmonE{}, with the black contours showing the \bsix{} continuum emission, and the orange contour showing the SiO line emission, highlighting the alignment with the blue-shifted cone.}
    \label{fig:mom0mom1}
\end{figure}
Originally identified by \citet{Kospal2021}, the continuum source \vmonE{} was re-detected in the \bthree{} and \bfour{} continuum data, yet remains undetected in the optical and NIR.
This implies that \vmonE{} is deeply embedded in dust that is optically thick at these wavelengths.
As discussed in Section~\ref{subsec:COemission}, a substantial portion of the $^{12}$CO emission lies in the foreground of \vmon{} and appears spatially anchored near \vmonE{}.
Fig.~\ref{fig:mom0mom1} shows the $^{12}$CO moment-1 map {where regions of negative moment-0 are masked.}
The velocity field is centered on \vmonE's velocity, with blue- and red-shifted lobes on opposite sides of its continuum emission.
The inset of Fig.~\ref{fig:mom0mom1} highlights the spatial alignment between the SiO jet and the blue-shifted cone of $^{12}$CO emission.
This alignment supports the interpretation of the observed $^{12}$CO feature as part of an outflow driven by \vmonE{}.
From $^{13}$CO we measure a central line of sight velocity of approximately $23.19\pm0.03$\kms (see Appendix~\ref{appendix:los}).
{We note that this value is subject to large systematic errors from contamination within the region that is not included in the presented uncertainty, so the resulting velocity estimate should be treated with caution.}
This velocity is blue-shifted by about $1$\kms{} with respect to \vmon{}.
This measurement sets as a lower limit for their relative motion and suggests that \vmonE{} is above the escape velocity of the system, hence probably not on a gravitationally bound orbit around \vmon{}.
Still, the close spatial and spectral proximity between \vmonE{} to \vmon{} suggests a dynamical association, further supported by a tentative connecting structure, faintly suggested in both \ceo{} and \ntdp{} moment-0 maps.

These observations raise the question of the dynamic history of the object driving the $^{12}$CO outflow.
The detection of \ceo{}, \dcop{}, and \ntdp{} is commonly interpreted as indicative of a protostellar envelope \citep{Tychoniec2021}.
Thus, \vmonE{} may represent a protostar forming from the same molecular reservoir that \vmon{} is accreting from \citep[similar to scenarios described by e.g.][]{Kuffmeier2019}, or it could be a stellar flyby.
In fact, the object \vmonSE{} presents another viable flyby candidate.
The line of sight velocity of \vmonSE{} is measured to be 23.55\kms{} from $^{12}$CO, so blue-shifted by 0.7\kms{} with respect to \vmon{} and likely above the escape velocity of the system.
Flybys have been shown to produce spiral features in circumstellar dust and gas and connecting structures between the flyby components \citep[e.g.,][]{Cuello2020}, suggesting that such a scenario may be at play here.
Interestingly, recent work by \citet{Fu2024} discussed the possibility that such a connecting feature induced by a flyby may gravitationally fragment to form future free-floating giant planets.

In terms of stellar accretion behavior, several studies pointed out that flybys may generate periods of enhanced accretion rates on the interacting stars, causing FUor events \citep[e.g.,][]{Forgan2010,Vorobyov2021,Borchert2022}. 
However, the simulated flybys produced sharply peaked bursts that are very unlike the observed extended classic FUor outbursts \citep[cf. section 5.4 in][for a discussion]{NayakshinEkbakyan2024}.
{To determine whether \vmonE{} or \vmonSE{} are bound to or unbound from the FUor, more accurate line of sight velocity measurements are needed. In addition, a deeper examination of the surrounding environment may reveal connecting structures that can inform on their dynamical relationship.}

\subsection{Ejection from circumstellar environment?}\label{subsec:ejection}
Finally, we want to discuss the scenario that some of the observed mm-sources represent compact objects ejected from the circumstellar environment of \vmon{} due to a stellar intruder, and the connection of this with the outbursting state of \vmon{}.

Recent work indicated that disk Thermal Instability \citep{Bell1994} in conjunction with thermal evaporation of massive planets formed by GI in the outer parts of protoplanetary disks presents an attractive scenario for FU\,Ori itself \citep{Nayakshin2023}, and potentially for several other observed FUor objects \citep{NayakshinEkbakyan2024,Nayakshin2024}.
The key to this scenario is a massive protoplanetary disk that is able to both hatch the fragments at $\sim 50$\,au and to also drive them inward into the inner $\sim0.1$\,au rapidly enough.
We note that GI fragment interactions with other fragments were shown to occasionally speed up the process of fragment migration from the outer into the inner disk \citep{Cha2011,Vorobyov2018}.
The mechanics of this process is simple.
The two bodies exchange angular momentum, with one of them gaining and the other losing angular momentum in a close interaction, the former thus gets onto an eccentric orbit with a much smaller peri-center distance, whereas the latter gets kicked to a larger orbit or ejected.
The more massive the interaction partner, the more pronouced is the dynamic effect.
In this regard, \citet{Vorobyov2017} have shown that close passages of stellar intruders may unbind GI-fragments from the disk, scattering them towards the star or ejecting them from the system.
Importantly, these fragments may be massive enough to be in the high-mass brown dwarf and even low-mass star regime (cf. their Table~1).
Previous 2D and 3D GI disk simulations showed prolific formation of such fragments \citep[e.g.][]{Vorobyov2006,Vorobyov2010,Cha2011}, but were unable to resolve the innermost region where FUor outbursts originate.
\citet{Nayakshin2023} showed that there in the inner disk Thermal Instability creates such a hot environment that these fragments may evaporate in a steady-state manner, providing an explanation for the bewilderingly long duration of FUor outbursts.

There is a potential connection of this scenario for FUor outbursts to the very disturbed and surprisingly large-scale gaseous environment of \vmon{}. 
There are several candidates for an intruder star in the immediate environment of \vmon{}. \citet{Caratti2015} reported a stellar object at a separation of about $227\,$mas from the primary, confirmed as a bright source in SPHERE $H$-band in \citet{Weber2023}.
Further, VLT/ERIS observations in $L'$-band ($\lambda_{\rm obs}=3.8\,$µm) detected a candidate companion at a distance of $0\farcs9$ to the southwest of \vmon{}, roughly co-located with the fragmented spiral arm observed in the \bsix{} continuum (Dasgupta et al. in prep.).
A potential intruder star may carry with it a part of the disk of the primary; and the ejected objects also tend to be surrounded by a gaseous disk.
As a result, shortly after passage, the system looks very disturbed, with one or more gaseous filaments connecting the primary with the intruder and the ejectees \citep[cf. Figs. 2 and 3 in][]{Vorobyov2017}.
The structure naturally has a scale much larger than that of a typical protoplanetary disk, extending well outside a thousand au.

Finally, even if the interaction between fragment and intruder failed to be sufficiently strong, flybys may still encourage faster GI-fragment destruction via more rapid disk migration.
As 3D simulations of disk flybys show \citep{Forgan2010,Vorobyov2021,Borchert2022}, the inner $\sim10$\,au of disks get sufficiently perturbed by the interaction to lead to higher accretion rates onto the primary star.
Although this effect itself is unlikely to produce a steady FUor-outburst-type light curve, as we argued above, the increased gas accretion rate in the inner $\sim10$\,au may promote much more rapid fragment migration, which could then evaporate in the inner disk as in the scenario of \citet{Nayakshin2023}.

Thus, an interaction with an intruder could simultaneously explain the formation of the southeastern elongated feature that is connecting \vmonSE{} and \vmon{}, and \vmon's FUor outburst.
This scenario could be further tested by future observations of the system.
In particular, constraints on the grain size in the southeastern elongated structure could help to differentiate between an infalling filament and the flyby ``bridge” scenarios, as the grains would be larger in the latter case.

\section{Conclusions}\label{sec:conclusions}
We presented ALMA, VLT/MUSE, and VLT/SPHERE observations of the outbursting star \vmon{}.
Our analysis places the 2014 FUor outburst \citep{Maehara2014} and the fragmenting spiral arm reported by \citet{Weber2023} into the broader environmental context:
\begin{itemize}
    \item We identified multiple compact objects emitting at millimeter wavelengths and found signs of direct interaction with \vmon{} for at least two of them, \vmonSE{} and \vmonE{}.
    \item From the millimeter spectral index, we infer that the ALMA continuum signal of those compact objects traces thermal emission emanating from the center of dense gas cores.
    \item The fragmenting spiral arm is not detected in \bfour{} or \bthree{}, also consistent with thermal emission at given sensitivities.
    \item \vmonE{} shows characteristics of a class-0 protostar and drives an outflow composed of a slower component traced by $^{12}$CO and a faster, more collimated jet identified through SiO emission.
    \item \vmonSE{} is the most prominent in mm continuum emission and we detect rotation in $^{12}$CO and $^{13}$CO line data, indicating a compact object at its center.
    \item MUSE and SPHERE imaging show a connection between \vmonSE{} and \vmon{}. We infer that the structure is situated behind \vmon{}, and from counterparts in $^{13}$CO channels we find that it is blue-shifted -- the material is moving towards \vmon{}, likely representing infall onto the system.
    \item This directly raises the question how this infall is connected to the outbursting state of the FUor \vmon{}, and the large-scale spirals observed in the NIR and the fragmenting spiral arm observed by ALMA \citep{Weber2023}.
\end{itemize}
These multi-wavelength observations suggest that \vmon{} is in a unique and transient evolutionary phase.
The system displays dynamic disruption across multiple scales—ranging from eruptive accretion onto the stellar surface, to gravitationally unstable spiral arms in dust and gas, and extending to large-scale ($\gtrsim 10{,}000$\,au) gravitational interactions with neighboring protostars.
This raises key questions about how processes at different scales interconnect and whether the observed scenario is representative of the general star and planet formation paradigm.

\begin{acknowledgments}
We thank the referee Ruobing Dong for a constructive report that helped to improve the quality of this work.
\acknPW{}
\acknYEMS{}
\acknSeba{}
\acknjames{}
\acknbaobab{}
\ackngarufi{}
\acknfcsm{}
\acknAgnes{}
\acknpuelche{}
\acknVLT{}
\acknalma
\end{acknowledgments}

\vspace{5mm}
\facilities{ALMA, VLT/MUSE, VLT/SPHERE}

\software{
This work has made use of the IRDAP-pipeline \citep[][]{vanHolstein2020} for the processing of SPHERE/IRDIS data,
the CASA software \citep[][]{CASA} for the processing of the ALMA data
and the {\tt mpdaf.obj}-package \citep{Bacon2016} for creating the MUSE cube.
We used {\tt bettermoments} \citep{bettermoments} for creating moment-0 and moment-1 maps for the molecular line emission in the ALMA data.
We used IPython \citep{ipython}, NumPy \citep{numpy} and Matplotlib \citep{Matplotlib} for data analysis and creating figures. 
}

\bibliography{main}{}
\bibliographystyle{aasjournal}

\appendix
\section{ALMA spectrum}\label{appendix:spectrum}
\begin{figure}
    \centering
    \includegraphics[width=\columnwidth]{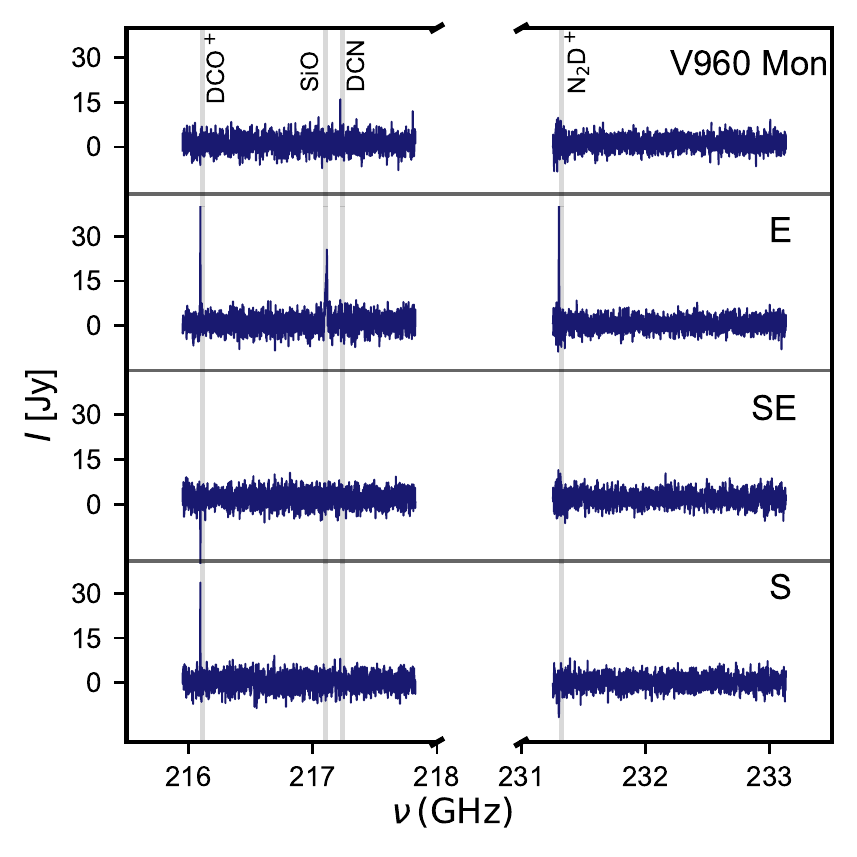}
    \caption{ALMA \bsix{} spectrum for four sources in the FOV. The vertical lines mark the rest emission frequencies of detected molecular transitions.}
    \label{fig:ALMA_spectra}
\end{figure}

We examined the \bsix{} continuum channels to identify any notable molecular emission. We focused on the continuum sources in the field of view, placing a circular aperture mask of radius \(1\farcs0\) on each source. By integrating the flux in each channel, we derived the spectra shown in Fig.~\ref{fig:ALMA_spectra} (before continuum subtraction).
The spectrum of \vmon{} is dominated by DCN.
The right panel of Fig.~\ref{fig:molecules} shows that this contribution stems from the fragmenting spiral arm inside the aperture centered on \vmon{}.
In \vmonE{}, three lines (\dcop{}, SiO, and \ntdp{}) are clearly present. \vmonSE{} displays no significant detections, although strong negative values near the \dcop{} line are attributed to contaminated baselines. For \vmonS{}, \dcop{} is the only pronounced feature.

\begin{figure*}
    \centering
    \includegraphics[]{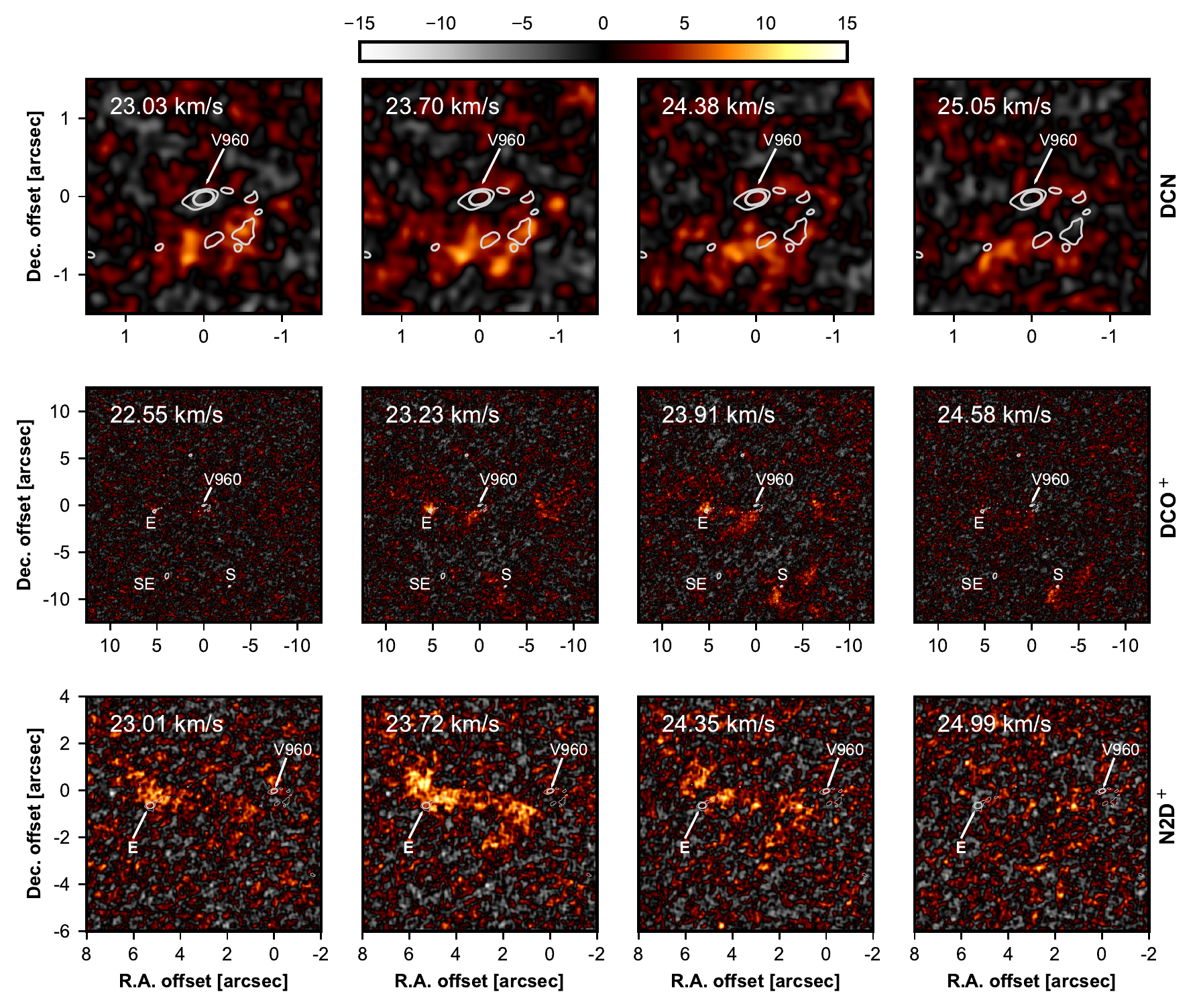}
    \caption{We show the intensity in units of mJy\,beam$^{-1}$ maps for the channels containing significant signal for the molecules DCN (top row), \dcop{} (center) and \ntdp{} (bottom row).
    For each of those molecular lines we found contributions limited to four successive channels.
    In each image, the white contours show the \bsix{} continuum flux at 5 and 10\,$\sigma_{\rm rms}$.
    The mm-sources in the FOV are marked with white labels.
    }
    \label{fig:molchannels}
\end{figure*}
Fig.~\ref{fig:molchannels} illustrates the individual channels containing notable flux for DCN, \dcop{}, and \ntdp{}. The fields of view vary to highlight different sources. Both DCN and \dcop{} channels reach an rms-noise level of $\sigma_{\rm rms}=1.8\,\mathrm{mJy}\,\mathrm{beam}^{-1}$, while \ntdp{} has $\sigma_{\rm rms}=2.5\,\mathrm{mJy}\,\mathrm{beam}^{-1}$. In the top row of Fig.~\ref{fig:molchannels}, DCN emission exceeding 3\,$\sigma_{\rm rms}$ appears only around the \bsix{} continuum clumps, outlined by white contours.
In the central row (\dcop{}), emission is strongest around \vmonE{} and \vmonS{}, with tentative flux in the area between \vmonE{} and \vmon{} that appears to connect to the fragmented continuum spiral arm.
This interpretation is further supported by \ntdp{} in the bottom row, which shows a similar alignment in those channels.
In both cases (\dcop{} and \ntdp{}) the emission from the area between \vmon{} and \vmonE{} is strongest for the channel that lies between the two objects systematic line of sight velocities.

\section{Line of sight Velocities}\label{appendix:los}
We aim to measure the line of sight velocities for the compact objects in the FOV from the Doppler shifts of CO molecular lines.
The contamination and low SNR of the optically thinner molecules complicate precise velocity measurements around the continuum sources.
We estimate the systematic velocity of \vmon{}, \vmonE{}, \vmonSE{} and \vmonS{}.
\vmonN{} does not show any associated molecular line emission, so is not included in this analysis.
\begin{figure*}
    \centering
    \includegraphics[width=\textwidth]{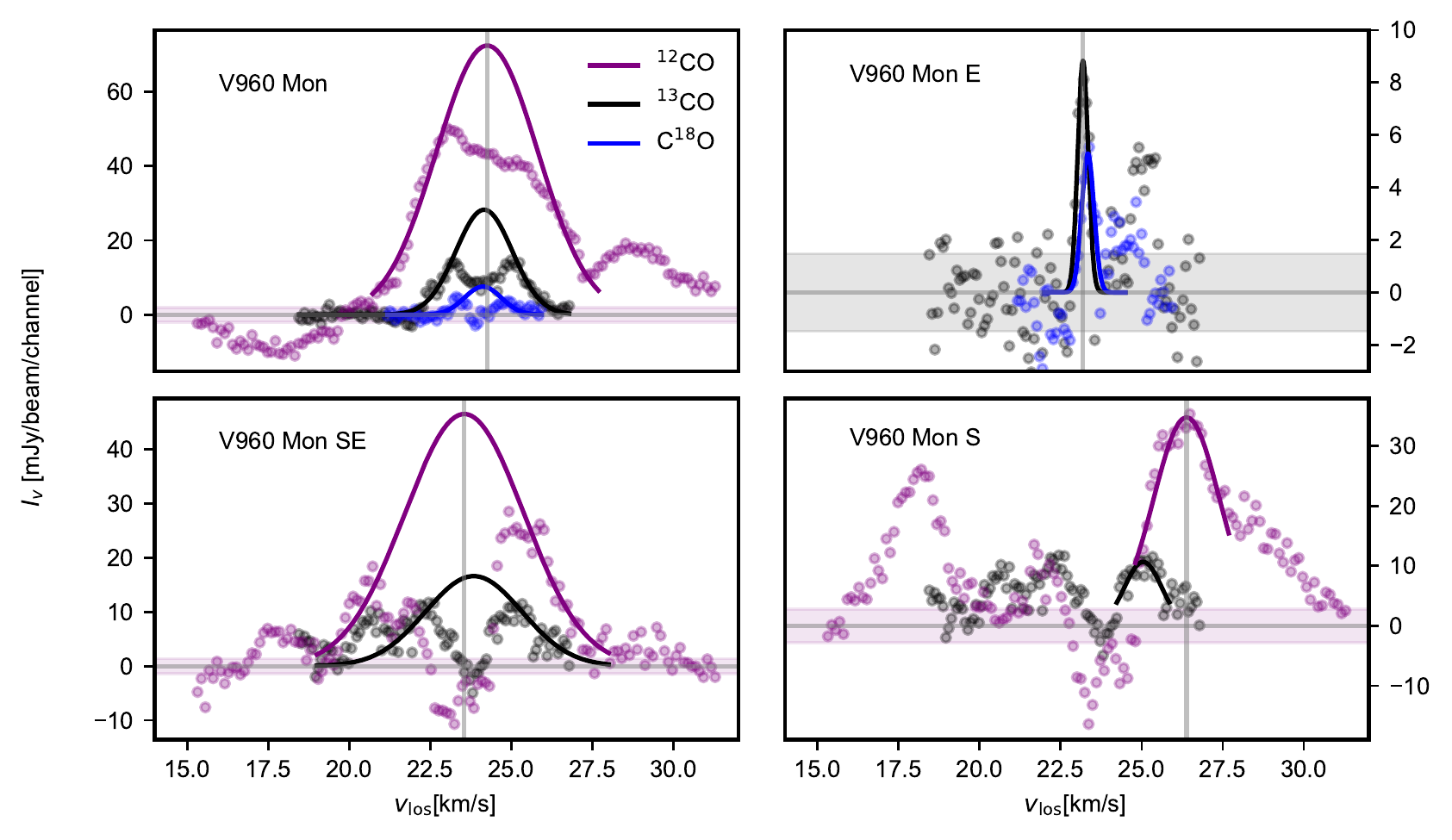}
    \caption{CO spectra from $1''$ around the four sources \vmon{}, E, SE and S, from the tracers $^{12}$CO, $^{13}$CO and C$^{18}$O (where applicable). The circles show the integrated flux in each channel, the lines correspond to Gaussian fits. We mark the selected representative line of sight velocity by a vertical line. {The vertical shaded area corresponds to the expected rms-noise of the spectrum. The is approximated by the value for the aperture-averaged flux from the $^{12}$CO, except for \vmonE{} where we use the value inferred from $^{13}$CO.}}
    \label{fig:vmon_co_shift}
\end{figure*}
For \vmon{}, the spectrum was extracted using a $1''$ aperture on the position of the star.
{We calculate the rms-noise of the spectrum from the aperture-averaged data.}
{The four panels of Fig.~\ref{fig:vmon_co_shift} show the spectra in dependence on the systematic velocity for four different objects.
The shaded area depicts the rms-noise between $-\sigma_{\rm rms}$ and $+\sigma_{\rm rms}$, for \vmon{}, \vmonSE{} and \vmonS{} corresponding to $^{12}$CO, for \vmonE{} corresponding to $^{13}$CO.}
For {\vmon{}, for} all three species, we fit a Gaussian to the wings of the emission profile.
The center of these Gaussians is very similar for the different isotoplogues, with $\bar{v}_{12} = (24.25\pm 0.03)$\kms{}, $\bar{v}_{13} = (24.16\pm0.03)$\kms{} and $\bar{v}_{18} = (24.12\pm0.09)$\kms, for $^{12}$CO, $^{13}$CO and \ceo, respectively. We choose the value of the $^{12}$CO measurement as a representative value, due to the higher SNR and the better velocity resolution.
The double-peak feature for $^{13}$CO and \ceo{} probably indicates of rotation around the primary.
However, we can not rule out that the diminution at the systematic velocity is due to self-absorption of the molecular line.

For \vmonE{}, the $^{12}$CO emission is strongly attenuated.
The SNR for $^{13}$CO and \ceo{} is low, but we detect a peak at the object's location (using a mask radius of $1\farcs0$) that is spectrally roughly consistent between the two isotopologues. The fit to the data points is shown in the second panel of Fig.~\ref{fig:vmon_co_shift}, with values of $\bar{v}^{\rm E}_{13} = (23.19\pm0.03)$\kms{} and $\bar{v}^{\rm E}_{18} = (23.34\pm0.04)$\kms.
We note that especially at higher velocities the CO that is co-moving with the primary seems to contaminate the spectrum.

The third panel of Fig.~\ref{fig:vmon_co_shift} shows the $^{12}$CO and $^{13}$CO spectrum for \vmonSE{}.
Here, we limit the fit to the area in which we detect significant \bsix{} continuum emission $I_{\rm 6}\geq 0.22$mJy\,beam$^{-1}$.
We do this because we detect rotation centered on the continuum in this area.
{From visual inspection of the $^{13}$CO moment-1 map (shown in the inset of the lower central panel of Fig.~\ref{fig:12CO}) that shows rotation around \vmonSE{}, we expect the central velocity of this rotation between 23 and 24\kms{}.
Also the emission around \vmonSE{} seems to be strongly attenuated at these velocities. We, therefore, fit the Gaussian only to the wings, which seem to be less affected by the contamination}.
Same as for \vmon{} we fit a Gaussian to the wings of the emission {from around \vmonSE{}}, and retrieve velocity centers of $\bar{v}^{\rm SE}_{12} = (23.55\pm 0.07)$\kms{} and $\bar{v}^{\rm SE}_{13} = (23.84\pm 0.14)$\kms.

Finally, the rightmost channel of Fig.~\ref{fig:vmon_co_shift} shows the spectrum from \vmonS{} using a mask of $0\farcs5$.
The $^{13}$CO emission shows a low SNR, the $^{12}$CO emission shows two peaks, one blue-shifted and one red-shifted with respect to \vmon{}. From inspecting the channels, we find that the blue shifted peak at around 18\kms is very extended over the entire southeastern area and thus probably due to cloud contamination.
The red-shifted peak, on the other hand, represents concentrated emission from around the continuum source.
We fit a Gaussian to the peak and measure a velocity of 26.38\kms.

\end{document}

%% file: acknowledgments.tex
\newcommand{\acknPW}{P.W. acknowledges support from FONDECYT grant 3220399.}
\newcommand{\acknSeba}{S.P. acknowledges support from FONDECYT Regular grant 1231663.}
\newcommand{\acknYEMS}{This work was funded by ANID -- Millennium Science Initiative Program -- Center Codes NCN2021\_080 and NCN2024\_001.}
\newcommand{\acknjames}{J.M. acknowledges support from FONDECYT de Postdoctorado 2024 \#3240612.}
\newcommand{\acknbaobab}{H.B.L. is supported by the National Science and Technology Council (NSTC) of Taiwan (Grant Nos. 111-2112-M-110-022-MY3, 113-2112-M-110-022-MY3).}
\newcommand{\ackngarufi}{The research activities described in this paper were carried out with contribution of the Next Generation EU funds within the National Recovery and Resilience Plan (PNRR), Mission 4 - Education and Research, Component 2 - From Research to Business (M4C2), Investment Line 3.1 - Strengthening and creation of Research Infrastructures, Project IR0000034 – “STILES - Strengthening the Italian Leadership in ELT and SKA”.}
\newcommand{\acknalma}{This paper makes use of the following ALMA data: ADS/JAO.ALMA\#2016.1.00209.S and ADS/JAO.ALMA\#2019.1.01144.S. ALMA is a partnership of ESO (representing its member states), NSF (USA) and NINS (Japan), together with NRC (Canada), NSTC and ASIAA (Taiwan), and KASI (Republic of Korea), in cooperation with the Republic of Chile. The Joint ALMA Observatory is operated by ESO, AUI/NRAO and NAOJ. The National Radio Astronomy Observatory is a facility of the National Science Foundation operated under cooperative agreement by Associated Universities, Inc.}
\newcommand{\acknpuelche}{This work made use of the Puelche cluster hosted at CIRAS/USACH.}
\newcommand{\acknVLT}{Based on observations collected at the European Southern Observatory under ESO programmes 098.C-0422(B) and 0106.C-0510(A).}
\newcommand{\acknAgnes}{This work was also supported by the NKFIH NKKP grant ADVANCED 149943 and the NKFIH excellence grant TKP2021-NKTA-64. Project no.149943 has been implemented with the support provided by the Ministry of Culture and Innovation of Hungary from the National Research, Development and Innovation Fund, financed under the NKKP ADVANCED funding scheme.}
\newcommand{\acknfcsm}{FCSM received financial support from the European Research Council (ERC) under the European Union’s Horizon 2020 research and innovation programme (ERC Starting Grant "Chemtrip", grant agreement No 949278).}